\documentclass[useAMS,usenatbib]{mn2e}
\usepackage{graphicx, amssymb}
\pdfoutput=1

\newcommand{\km}{${\rm km\,s}^{-1}$}
\newcommand\nodata{ ~$\cdots$~ }%
\newcommand{\hst}{{\em HST}}

\newcommand{\vlsr}{$v_{\rm LSR}$\relax}

\def\lesssim{\mathrel{\hbox{\rlap{\hbox{%
 \lower4pt\hbox{$\sim$}}}\hbox{$<$}}}}
\def\gtrsim{\mathrel{\hbox{\rlap{\hbox{%
 \lower4pt\hbox{$\sim$}}}\hbox{$>$}}}}
\let\la=\lesssim                
\let\ga=\gtrsim

\newcommand{\hi}{H$\;${\small\rm I}\relax}
\newcommand{\hii}{H$\;${\small\rm II}\relax}

\newcommand{\alii}{Al$\;${\small\rm II}\relax}
\newcommand{\aliii}{Al$\;${\small\rm III}\relax}

\newcommand{\nni}{N$\;${\small\rm I}\relax}

\newcommand{\caii}{Ca$\;${\small\rm II}\relax}

\newcommand{\cii}{C$\;${\small\rm II}\relax}
\newcommand{\ciii}{C$\;${\small\rm III}\relax}
\newcommand{\civ}{C$\;${\small\rm IV}\relax}

\newcommand{\oi}{O$\;${\small\rm I}\relax}

\newcommand{\ovi}{O$\;${\small\rm VI}\relax}

\newcommand{\sii}{S$\;${\small\rm II}\relax}
\newcommand{\siii}{Si$\;${\small\rm II}\relax}
\newcommand{\siiii}{Si$\;${\small\rm III}\relax}
\newcommand{\sixii}{Si$\;${\small\rm XII}\relax}
\newcommand{\siiv}{Si$\;${\small\rm IV}\relax}

\newcommand{\mgii}{Mg$\;${\small\rm II}\relax}
\newcommand{\mgx}{Mg$\;${\small\rm X}\relax}

\newcommand{\feii}{Fe$\;${\small\rm II}\relax}

\newcommand{\neviii}{Ne$\;${\small\rm VIII}\relax}

\newcommand{\ciis}{C$\;${\scriptsize\rm II}\relax}
\newcommand{\civs}{C$\;${\scriptsize\rm IV}\relax}

\newcommand{\siiis}{Si$\;${\scriptsize\rm II}\relax}
\newcommand{\siiiis}{Si$\;${\scriptsize\rm III}\relax}

\title[HVCs: Streams of gas in the Milky Way halo]{High-velocity clouds as streams of ionized and neutral gas in the halo of the Milky Way\thanks{Based on observations made with the NASA/ESA Hubble Space Telescope, obtained at the Space Telescope Science Institute, which is operated by the Association of Universities for Research in Astronomy, Inc. under NASA contract No. NAS5-26555.}}
\author[N. Lehner et al.]{N. Lehner$^{1}$\thanks{e-mail:N.Lehner@nd.edu}, J.~C. Howk$^{1}$, C. Thom$^{2}$, A.~J. Fox$^{2}$, J. Tumlinson$^{2}$, T.~M. Tripp$^{3}$, and \newauthor J.~D. Meiring$^{3}$ \\
$^{1}$Department of Physics, University of Notre Dame,  Notre Dame, IN 46556, USA\\
$^{2}$Space Telescope Science Institute, Baltimore, MD 21218, USA \\
$^{3}$Department of Astronomy, University of Massachusetts, Amherst, MA 01003, USA}

\begin{document}

\date{Accepted XXX. Received XXX; in original form XXX}

\pagerange{\pageref{firstpage}--\pageref{lastpage}} \pubyear{2011}

\maketitle
 \label{firstpage}

\begin{abstract}
High-velocity clouds (HVC), fast-moving ionized and neutral gas clouds found at high galactic latitudes, may play an important role in the evolution of the Milky Way.  The extent of this role depends sensitively on their distances and total sky covering factor.  We search for HVC absorption in {\it Hubble Space Telescope}\ high resolution ultraviolet spectra of a carefully selected sample of 133 AGN using a range of atomic species in different ionization stages (e.g., \oi, \cii, \civ, \siii, \siiii, \siiv). This allows us to identify neutral, weakly ionized, or highly ionized HVCs over several decades in \hi\ column densities. The sky covering factor of UV-selected HVCs with $|v_{\rm LSR}|\ge 90$ \km\  is $68\% \pm 4\%$ for the entire Galactic sky. About $74\%$ of the HVC directions have $N($\hi$)<3\times 10^{18}$ cm$^{-2}$ and 46\%  have $N($\hi$)<8\times 10^{17}$ cm$^{-2}$. We show that our survey is essentially complete, i.e., an undetected population of HVCs with extremely low H (\hi+\hii) column density is unlikely to be important for the HVC mass budget. We confirm that the predominantly ionized HVCs contain at least as much mass as the traditional \hi\ HVCs and show that large \hi\ HVC complexes have  generally ionized envelopes extending far from the \hi\ contours. There are also large regions of the Galactic sky that are covered with ionized high-velocity gas with little \hi\ emission nearby. We show that the covering factors of HVCs with $90 \le |v_{\rm LSR}|\la 170$ \km\ drawn from the AGN and stellar samples are similar. This confirms that these HVCs are within 5--15 kpc of the sun. The covering factor of these HVCs drops with decreasing vertical height, which is consistent with HVCs being decelerated or disrupted as they fall to the Milky Way disk. The HVCs with $|v_{\rm LSR}|\ga 170$ \km\ are largely associated with the Magellanic Stream at $b<0\degr$ and its leading arm at $b>0\degr$ as well as other large known \hi\ complexes. Therefore there is no evidence in the Local Group that any galaxy shows  a population of HVCs extending much farther away than 50 kpc from its host, except possibly for those tracing remnants of galaxy interaction.
\end{abstract}


\section{Introduction}
The ability of galaxies to form stars depends sensitively on the content and physical conditions of their gas, which in turn is dictated by internal effects, such as feedback, as well as by the external interaction of galaxies with their surroundings. A galaxy may exchange mass with its circumgalactic environment through the infall of intergalactic matter or of remnants resulting from galaxy interactions, and through the outflows driven by stellar and AGN feedback. Despite many theoretical and observational efforts \citep[a non exhaustive list of references includes][]{oort70,steidel92,keres05,bouche07,peek08,dekel09,prochaska09,oppenheimer10,fumagalli11,marinacci11,tripp11,tumlinson11}, the cycle of baryons in and out of galaxies and its implications for galaxy evolution are not yet fully understood. This exchange of baryons between galaxies and their environments must, however, result in a net gain of mass if galaxies are to form stars over many billions of years.

In our own Milky Way, it has been long known that there must be streams or flows of gas through the Galactic halo. Signatures of these flows are revealed by the  high-velocity clouds (HVCs), clouds  moving in the local standard of rest (LSR) frame at $|\mbox{\vlsr}| \ge 90$ \km\ \citep{wakker97}. The unknown distances to these clouds for a long time meant we did not know if the HVCs were associated with streams occurring near the Milky Way \citep[e.g.,][]{oort70} rather than the IGM of the Local Group \citep[e.g.,][]{blitz99,gnat04}. Furthermore, the distances to the HVCs are required for quantifying their basic physical properties (several scale with the distance, e.g., the mass $M\propto d^2$). Since the last large summary on the distances and metallicities of the \hi\ HVCs published 10 years ago \citep{wakker01}, our knowledge of these clouds has dramatically improved. Distances of several individual large HVC complexes seen in \hi\ emission are now determined, placing them at heliocentric distances of about 4--15 kpc \citep[e.g.,][]{ryans97,wakker01,thom06,thom08,wakker07,wakker08,lehner10,smoker11}. Among the large complexes (in terms of mass and solid angle), only the Magellanic Stream \citep[e.g.,][]{putman98,bruns05,nidever08} is much farther away, possibly extending to 80--200 kpc according to recent numerical models of the interaction of the Large and Small Magellanic Clouds \citep{besla12}. 

With the advent of sensitive high resolution UV spectrographs including the {\it Far Ultraviolet Spectroscopic Explorer (FUSE)}\ and those of the {\it Hubble Space Telescope (HST)}, it has also become obvious that HVCs are both neutral and ionized in view of the high-velocity absorption detected  in resonance lines  of neutral, weakly ionized, and highly ionized species in the spectra of AGN \citep[e.g.,][]{sembach95,sembach99,sembach00,savage00,lehner01,sembach03,wakker03,collins05,fox06,shull09,richter09,tripp12}. The importance of ionized gas can be demonstrated simply by considering the fraction of the sky covered by HVCs at \hi\ column sensitivities.  While only $18\%$ is covered at a sensitivity of $\log N($\hi$) \ga 18.5$ \citep{wakker91}, a limit easily detected in \hi\ 21 cm observations, deeper \hi\ emission observations ($\log N($\hi$)\ga 17.9$) yield a factor larger covering factor, $37\%$ \citep{murphy95,lockman02}. A much larger fraction of the sky is even covered by the HVCs seen in absorption, which are sensitive to much lower $N($\hi$)$: for the HVCs with \ovi\ absorption, the HVC covering factor is about 60--75\% \citep{sembach03,fox06}, and a high covering factor was also found for the HVCs detected via the strong line of \siiii\ \citep[][hereafter CSG09]{shull09,collins09}. 

Many of these HVCs observed in absorption have been often dubbed as ionized HVCs or highly ionized HVCs since in many cases a large fraction of the gas is photoionized or collisionally ionized, i.e., $N($\hii$)\gg N($\hi$)$. In this work, we simply label them as HVCs. To differentiate them from the HVCs  observed in \hi\ emission, we refer to the latter category as \hi\ HVCs. We emphasize that the only real distinction with this definition is that the \hi\ HVCs are detectable via 21-cm emission. Thus they have a clear \hi\ sensitivity cut-off determined by the depth of the observations and limitation of radio emission studies. On the other hand, for the HVCs (i.e., those seen in UV absorption), the amount of \hi\ can vary greatly as demonstrated by the observed range in the \hi\ column density: $N($\hi$)\la 10^{15}$  cm$^{-2}$ to $N($\hi$)\ga 10^{20}$  cm$^{-2}$ derived from Lyman series absorption \citep{fox06,zech08} and \hi\ 21-cm emission observations toward the AGN; virtually {\it any}\ $N($\hi$)$ column densities are observed in the UV-selected HVC. However, even if the \hi\ column density can be small, the total H (\hi+\hii) column density of the  HVCs can still be substantial,  $\ga 10^{17}$--$10^{18}\, (Z/Z_{\odot})^{-1}$ cm$^{-2}$ on average \citep[e.g.,][]{sembach03,shull09}. As for the \hi\ HVCs, direct distance constraints are required for determining the masses of the HVCs seen in UV absorption and for characterizing their role in the evolution of the Milky Way and their relationship with the \hi\ HVC complexes. If they are in the Milky Way within several kpc, the HVCs seen in UV absorption are not only more important in number and mass, but their relative proximity may allow them  to  descend into the disk to fuel new generations of star formation \citep[e.g.,][]{shull09,lehner11}.

\citet[][hereafter LH11]{lehner11}  ruled out whole classes of HVCs using a novel and powerful method to statistically constrain the distance of the entire HVC population rather than just a few individual clouds. This was done by comparing the UV absorption  line detection rates of HVCs in COS and STIS spectra of stars (distance limited) and AGN (no distance limit).  They selected 28 stars (hereafter, the LH11 stellar sample) to be at the largest practical distances in order to distinguish between Milky Way halo and Local Group origins for the HVCs, resulting in targets at $|z|\ga 3 $ kpc from the Galactic plane and heliocentric distances $3 \le d \la 32$ kpc (although most of the stars are within $\langle d \rangle = 11.6 \pm 6.9$ kpc).  The stellar sightlines were carefully selected to be well distributed across the high latitude Galactic sky and avoid biases that {\it a priori}\ favor or disfavor the presence of HVCs along each sightline. They identified HVC absorption in the stellar spectra using the strongest lines of atomic, singly, and doubly ionized species available in the UV bandwidth (e.g., \cii\ $\lambda$1334, \civ\ $\lambda$1548, \siii\ $\lambda$1260, \siiii\ $\lambda$1206, \oi\ $\lambda$1302), most showing high [\siii/\oi] ratio, consistent with these HVCs being mostly ionized. The HVC sky covering factors of the HVCs with $90 \le |v_{\rm LSR}|\la 170$ \km\ determined from the stellar and AGN samples are $f_c \simeq 0.50$ and $\la 0.65 $, respectively (LH11). Unfortunately the HVC covering covering factor for the AGN sample  suffers from poorly understood selection criteria, resulting only in an upper limit on $f_c$. Nevertheless the similarity in the  covering factors between the distance limited sample and the sample with no distance limit imply that most of the HVCs  with $90 \le |v_{\rm LSR}|\la 170$ \km\ are within heliocentric distances $5$--$15$ kpc  (no HVC at $|v_{\rm LSR}|\ga 200$ \km\ was found in the stellar sample). Given the importance of HVC distances and the importance of the covering factor in determining $d$, one of our goals here is to robustly determine the covering factor of the HVCs from a well understood sample of AGN that can be directly compared to the LH11 stellar sample. To search for the HVCs, we do not focus on a single ion, but on a suite of atoms and ions (e.g., \oi, \cii, \civ, \siii, \siiii, \siiv, \alii) following the same method used by LH11. This allows us not only to securely detect HVC absorption, but also, as we will show, to sensitively search for HVCs at all levels of ionization. 

Here we present the results of our new study of HVCs, with the following organization.  In \S\ref{s-obs} we describe our samples of AGN, the data and their sensitivity, and the method for determining the presence of HVCs. In \S\ref{s-cov} we present our findings for the covering factors of the HVCs and VHVCs. With this new sample of AGN, we strengthen the LH11 results, and we extend their results  by using stars at smaller distances to study the dependence between the covering factor and the distance. With our much larger sample of AGN and much better sky coverage than the sample used in LH11, we are also able to gain insight into the origins and distances of the very high-velocity clouds (VHVCs, $|\mbox{\vlsr}| \ga 170$ \km).  In \S\ref{s-dist} we connect the HVCs detected in \hi\ emission and UV absorption based on their closed proximity on the sky, in velocity, and in space (distance). We discuss some of the implications of our results for the mass of the HVCs and their origins and fate in \S\ref{s-imp}. Finally, we present a summary in \S\ref{s-con}.

\begin{figure*}
  \includegraphics[width=16 truecm]{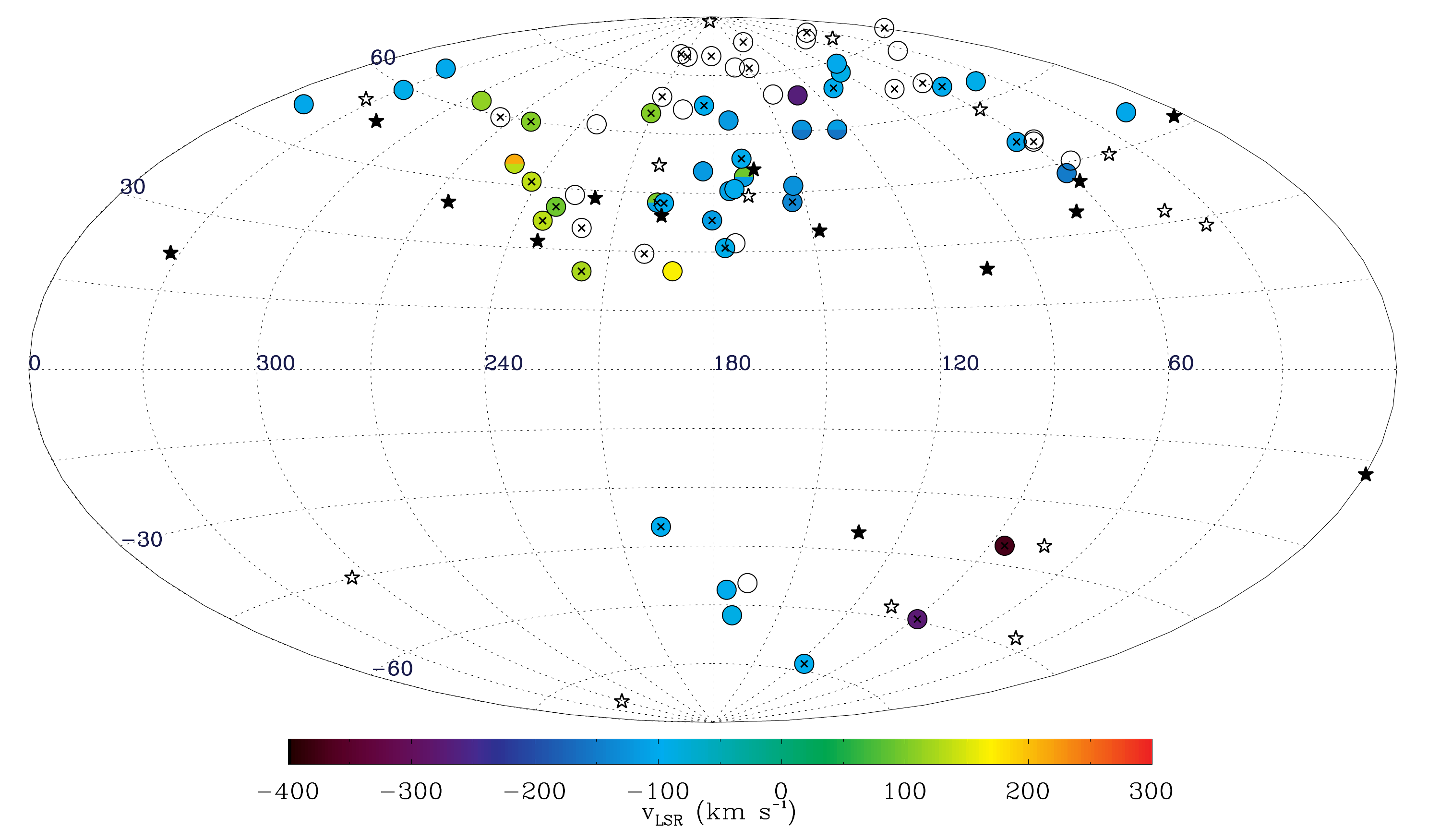}
  \caption{ Aitoff projection map of the survey directions for the TT sample. Sightlines are plotted in Galactic coordinates with longitude increasing from right to left.  A colored circle indicates an HVC along the line of sight while open circle implies no HVC along the sightline. Circles with a cross indicate that the sightline is part of the high-sensitivity sample with $10 \le W_{\rm lim} \le 42$ m\AA. The velocity value is color coded following the horizontal color bar. The star symbol shows the positions of the stars from the sample of LH11 where a filled symbol indicates that an HVC is detected in the foreground of the stars, while an open symbol shows the absence of HVC along the stellar sightline.
  \label{f-map}}
\end{figure*}

\section{AGN sample and data}\label{s-obs}

\subsection{Description of the main AGN sample}\label{s-redux}
The 67 AGN in our main sample were drawn from 3 \hst/COS large programs: 11598 (``How Galaxies Acquire their Gas: A Map of Multiphase Accretion and Feedback in Gaseous Galaxy Halos", PI: Tumlinson), 11741 (``Probing Warm-Hot Intergalactic Gas at $0.5 < z < 1.3$ with a Blind Survey for \ovi, \neviii, \mgx, and \sixii\ Absorption Systems, PI: Tripp), and 12248 (``How Dwarf Galaxies Got That Way: Mapping Multiphase Gaseous Halos and Galactic Winds Below $L^*$", PI: Tumlinson).\footnote{The program 12248 is not completed yet; only data taken prior to November 2011 are in this paper.} The aims of these 3 programs are to target science beyond the Milky Way halo. Therefore the choice of the targets was completely random with respect to the locations of the \hi\ HVCs. Hereafter we refer to this sample as the TT (Tripp--Tumlinson) sample. 

Information on the design and performance of COS can be found in \citet{green12}.  The data reduction and co-addition procedures are described in \citet{thom11} and \citet{meiring11}, and we refer the reader to these papers for more details on data processing. Notably the individual exposures were co-added in photon counts, rather than flux calibrated exposures. The shifts between the individual exposures were determined using the Galactic interstellar absorption lines.  The exposures were shifted to a common reference and co-added. All the data were then shifted to the local standard of rest (LSR) frame. A visual comparison between the Galactic \hi\ emission spectra from the Leiden/Argentine/Bonn (LAB) survey \citep{kalberla05} and the Galactic absorption indicates that the spectra were correctly shifted in the absolute LSR frame.  The COS resolution of the G130M and G160M gratings is $R\approx 17,000$, which is adequate for identifying HVCs \citep[][LH11]{lehner10}. The typical signal-to-noise of the co-added spectra are  about 4--9, $>20$, and 9--13 per resolution element for programs 11598, 11741, and 12248, respectively. A few sightlines were not considered in the programs 11598 and 12248 owing to the poor signal-to-noise of their spectra ($\la 3$) or excessive contamination from higher redshift absorbers. 

In Fig.~\ref{f-map}, we show  the sky distribution of the 67 TT targets, and in Table~\ref{t-sample} we list their Galactic coordinates. Most of the targets are situated in the northern Galactic hemisphere as the two largest samples (11598 and 12248) were selected from the Sloan Digital Sky Survey (SDSS). There is an excellent coverage of the Galactic sky in longitude and $b\ga 30\degr$. There is also good overall agreement in the sky coverage between the TT sample and sample of stars assembled by LH11 as demonstrated in Fig.~\ref{f-map}. 

\begin{table}
 \centering
 \begin{minipage}{7.5cm}
\scriptsize
  \caption{The AGN sample \label{t-sample}}
  \begin{tabular}{lccccc}
  \hline
   Name     &   $l$         &  $b$  & $v^{\rm HVC}_{\rm LSR}$ & $W_{\rm lim}$$^a$ & P$^b$\\
            &   (\degr)      & (\degr)& (\km)		& (m\AA)  &	\\
 \hline
          J1342+1844   & $     0.13 $ & $     +75.48 $  & \nodata	  & $  47.7   $ &  2  \\
          J1524+0958   & $    14.89 $ & $     +50.12 $  & $-106 	$ & $  19.1   $ &  3  \\
          J1409+2618   & $    34.67 $ & $     +72.59 $  & \nodata	  & $  10.0   $ &  3  \\
          J1451+2709   & $    39.62 $ & $     +63.43 $  & $ -95:	$ & $  41.4   $ &  2  \\
          J1330+2813   & $    42.37 $ & $     +81.23 $  & \nodata	  & $  42.1   $ &  1  \\
          J1619+3342   & $    54.59 $ & $     +45.18 $  & \nodata	  & $  30.7   $ &  1  \\
          J1445+3428   & $    56.74 $ & $     +64.59 $  & $-101 	$ & $  47.9   $ &  1  \\
          J1553+3548   & $    57.26 $ & $     +50.67 $  & \nodata	  & $  40.4   $ &  1  \\
          J1555+3628   & $    58.32 $ & $     +50.27 $  & \nodata	  & $  49.7   $ &  1  \\
          J1632+3737   & $    60.34 $ & $     +42.94 $  & $-152 	$ & $	9.5   $ &  3  \\
          J1330+3119   & $    61.27 $ & $     +80.41 $  & \nodata	  & $  36.7   $ &  2  \\
          J1435+3604   & $    61.52 $ & $     +66.26 $  & \nodata	  & $  60.2   $ &  1  \\
          J1550+4001   & $    63.96 $ & $     +50.94 $  & $-108 	$ & $  81.1   $ &  1  \\
          J1419+4207   & $    78.58 $ & $     +66.66 $  & \nodata	  & $  52.8   $ &  1  \\
          J1342+3829   & $    83.11 $ & $     +74.40 $  & $-102$	  & $  27.3   $ &  2  \\
          J2257+1340   & $    85.28 $ & $     -40.73 $  & $-370 	 $ & $  64.6   $ &  1  \\
          J2345--0059  & $    88.79 $ & $     -59.39 $  & $-276 	 $ & $  48.5   $ &  1  \\
          J1341+4123   & $    90.59 $ & $     +72.48 $  & $ -93 	 $ & $  27.0   $ &  3  \\
          J1322+4645   & $   107.71 $ & $     +69.44 $  & $-100:	 $ & $  47.0   $ &  1  \\
          J0042--1037  & $   115.13 $ & $     -73.36 $  & $ -98 	 $ & $  53.4   $ &  1  \\
          J1241+5721   & $   125.48 $ & $     +59.73 $  & $-158, -113	 $ & $  37.0   $ &  1  \\
          J1233+4758   & $   131.24 $ & $     +68.87 $  & $-270 	 $ & $  37.7   $ &  1  \\
          J1245+3356   & $   133.72 $ & $     +83.06 $  & \nodata	  & $  48.2   $ &  1  \\
          J1151+5437   & $   140.61 $ & $     +60.39 $  & $-155,-126:	$ & $  12.0   $ &  3  \\
          J1208+4540   & $   144.63 $ & $     +69.62 $  & \nodata	  & $  14.4   $ &  3  \\
          J1220+3853   & $   149.71 $ & $     +76.59 $  & \nodata	  & $  61.5   $ &  1  \\
          J1001+5944   & $   152.44 $ & $     +46.53 $  & $-128 	$ & $  31.1   $ &  2  \\
          J0928+6025   & $   154.10 $ & $     +42.44 $  & $-141 	$ & $  70.6   $ &  1  \\
          J1211+3657   & $   161.27 $ & $     +76.95 $  & \nodata	  & $  36.0   $ &  2  \\
          J0226+0015   & $   166.57 $ & $     -54.38 $  & \nodata	  & $  37.1   $ &  1  \\
          J0950+4831   & $   168.96 $ & $     +49.10 $  & $-116,  +92	$ & $  40.0   $ &  1  \\
          J1016+4706   & $   169.03 $ & $     +53.74 $  & $-100 	$ & $  43.9   $ &  1  \\
          J0212--0737  & $   171.06 $ & $     -62.64 $  & $ -91:	$ & $  29.2   $ &  2  \\
          J1103+4141   & $   172.52 $ & $     +63.52 $  & $-120 	$ & $  44.2   $ &  2  \\
          J0929+4644   & $   172.65 $ & $     +46.02 $  & $-100:	$ & $  25.9   $ &  2  \\
          J0809+4619   & $   173.33 $ & $     +32.21 $  & \nodata	  & $  31.4   $ &  2  \\
          J0925+4535   & $   174.42 $ & $     +45.54 $  & $-108 	$ & $  29.9   $ &  2  \\
          J0235--0402  & $   174.46 $ & $     -56.16 $  & $-100 	$ & $  23.3   $ &  3  \\
          J0803+4332   & $   176.40 $ & $     +31.00 $  & $-95^c$	  & $  57.2   $ &  1  \\
          J0843+4117   & $   180.21 $ & $     +38.06 $  & $-114 	$ & $  59.2   $ &  2  \\
          J1210+3157   & $   181.58 $ & $     +79.96 $  & \nodata	  & $  55.9   $ &  2  \\
          J0949+3902   & $   183.55 $ & $     +50.57 $  & $-120 	$ & $  37.9   $ &  2  \\
          J1112+3539   & $   184.71 $ & $     +67.34 $  & $-99^c$	  & $  59.5   $ &  1  \\
          J0751+2919   & $   191.34 $ & $     +25.05 $  & $+175 	$ & $  13.5   $ &  3  \\
          J1104+3141   & $   195.45 $ & $     +66.24 $  & \nodata	  & $  41.1   $ &  2  \\
          J0912+2957   & $   195.86 $ & $     +42.34 $  & $-100:	$ & $  71.5   $ &  2  \\
          J0401--0540  & $   196.42 $ & $     -39.97 $  & $ -99:	$ & $  52.8   $ &  1  \\
          J0914+2823   & $   198.13 $ & $     +42.45 $  & $-110:, +95:  $ & $  47.2   $ &  1  \\
          J0820+2334   & $   199.80 $ & $     +29.42 $  & \nodata	  & $  42.0   $ &  1  \\
          J1204+2754   & $   206.11 $ & $     +79.62 $  & \nodata	  & $  45.4   $ &  2  \\
          J1117+2634   & $   209.19 $ & $     +69.21 $  & \nodata	  & $  45.9   $ &  2  \\
          J1059+2517   & $   210.82 $ & $     +64.96 $  & $+102^c$	  & $  46.0   $ &  2  \\
          J1207+2624   & $   214.56 $ & $     +80.11 $  & \nodata	  & $  51.4   $ &  2  \\
          J0826+0742   & $   216.89 $ & $     +24.66 $  & $+125 	$ & $  46.5   $ &  2  \\
          J0910+1014   & $   219.82 $ & $     +35.48 $  & \nodata	  & $  94.0   $ &  1  \\
          J0947+1005   & $   225.59 $ & $     +43.59 $  & \nodata	  & $  36.4   $ &  2  \\
          J0943+0531   & $   230.04 $ & $     +40.41 $  & $ +90 	$ & $  55.9   $ &  1  \\
          J0935+0204   & $   232.39 $ & $     +36.84 $  & $+132 	$ & $  55.4   $ &  1  \\
          J1059+1441   & $   232.72 $ & $     +61.19 $  & \nodata	  & $  37.2   $ &  2  \\
          J1022+0132   & $   242.16 $ & $     +46.07 $  & $+135 	$ & $  55.3   $ &  1  \\
          J1051--0051  & $   252.24 $ & $     +49.88 $  & $+131, +210	$ & $  37.8   $ &  2  \\
          J1133+0327   & $   261.35 $ & $     +59.87 $  & $+102 	$ & $  52.3   $ &  1  \\
          J1157--0022  & $   275.71 $ & $     +59.64 $  & \nodata	  & $  49.2   $ &  1  \\
          J1233--0031  & $   293.11 $ & $     +61.99 $  & $+130:^c$	   & $  29.0   $ &  1  \\
          J1342--0053  & $   328.82 $ & $     +59.37 $  & $ -95 	$ & $  35.6   $ &  1  \\
          J1342+0505   & $   333.75 $ & $     +64.84 $  & $-103^c$	  & $  38.0   $ &  2  \\
          J1437--0147  & $   348.72 $ & $     +51.37 $  & $-107 	$ & $  15.4   $ &  3  \\
\hline
\end{tabular}
Note: Absence of  value in the velocity column indicates that there is no evidence of HVC along the line of sight.   Uncertainty in the velocity is dominated by the COS wavelength calibrations, yielding an error of $\sim$10 \km; the statistical error is generally $<3$ \km, except for those marked with a colon. A velocity value followed by a colon is uncertain owing to blending with the lower velocity gas.
\\
($a$): $3\sigma$ equivalent width detection limits. We estimated $W_{\rm lim}$ near the \siiiis\ absorption, but the distribution of $W_{\rm lim}$ for \ciis\ or \siiis\ would be very similar (although the actual value of $W_{\rm lim}$ of \ciis\ or \siiis\ may be different for a given sightline; see \S\ref{s-sens}).
($b$): \hst\ Program: (1) 11598, (2) 12248, (3)  11741. ($c$) HVC only detected in \siiiis\ and \civs\ (not \siiis\ or \ciis). 
\end{minipage}
\end{table}

\subsection{Additional sample of AGN}
In order to achieve the best sky coverage possible, we also consider two additional samples: 1) 18 publicly available COS AGN spectra from the Guaranteed Time Observer (GTO) IGM program, and 2) 49 AGNs drawn from the STIS E140M and G140M AGN sample from CSG09. 

The GTO COS spectra were reduced following the procedure described in \S\ref{s-redux}. We only considered AGN that were initially targeted for IGM science, not the AGN from the GTO Galactic--HVC program to avoid obvious selection bias. The GTO sample is summarized in Table~\ref{t-gto}. 

\begin{table} 
\centering
 \begin{minipage}{6cm}
\scriptsize
  \caption{The GTO IGM sample \label{t-gto}}
  \begin{tabular}{lccc}
  \hline
   Name     &   $l$         &  $b$  & $v^{\rm HVC}_{\rm LSR}$ \\
            &   (\degr)      & (\degr)& (\km)		\\
 \hline
           1ES1553+113      & $     21.91  $ & $     +43.96  $  &   \nodata	     \\
           PG1259+593	    & $    120.56  $ & $     +58.05  $  &   $-142	$    \\
           SBS1122+594      & $    141.80  $ & $     +54.71  $  &    $-95	$    \\
            VIIZW244	    & $    136.66  $ & $     +32.67  $  &    $-93	$    \\
           3C263	    & $    134.16  $ & $     +49.74  $  &   $-166,   -122$   \\
           PG1115+407	    & $    172.23  $ & $     +66.67  $  &    $-110:	 $   \\
           MRK421	    & $    179.83  $ & $     +65.03  $  &   $-119	$    \\
           HS1102+3441      & $    188.56  $ & $     +66.22  $  &   \nodata	     \\
           TON580	    & $    194.94  $ & $     +72.03  $  &   \nodata	     \\
           HE0238--1904     & $    200.48  $ & $     -63.63  $  &   \nodata	    \\
           PKS0405--123     & $    204.93  $ & $     -41.76  $  &   $+110      $    \\
           HE0226--4110     & $    253.94  $ & $     -65.78  $  &   $+156,  +208$    \\
           HE0435--5304     & $    261.02  $ & $     -41.38  $  &   $+100:     $    \\
           HE0439--5254     & $    260.69  $ & $     -40.90  $  &   $+106,  +300$    \\
           RXJ0439.6--5311  & $    261.22  $ & $     -40.93  $  &   $+116,  +300$    \\
           HE0153--4520     & $    271.80  $ & $     -67.98  $  &   $+110,  +197$    \\
           PKS2005--489     & $    350.37  $ & $     -32.60  $  &   \nodata	    \\
           RXJ2154.1--4414  & $    355.18  $ & $     -50.87  $  &   \nodata	    \\
\hline
\end{tabular}
Note: A velocity value followed by a colon is uncertain owing to blending with the lower velocity gas.
\end{minipage}
\end{table}

The only explicit two criteria for selecting sightlines in CSG09 were that 1) S/N\,$\ge 3$, and 2) little apparent contamination from the QSO absorbers. However, we need to remove any potential bias from their sample, as, e.g., some of their AGN were initially targeted for HVC science (with a priori knowledge from past observations that an HVC was along the line of sight). In order to identify AGN originally targeted for HVC science, we reviewed the MAST program abstract and title for each target in CSG09 and found that 7 sightlines were explicitly selected to study HVCs. Those were not considered further. From the appendix in CSG09, we determined that 6 lines of sight  had an HVC detection solely based on \siiii\ absorption (for the other HVCs, they have confirmation of the absorption using several species and/or transitions, i.e., consistent with our criteria from \S\ref{s-search}). This differs from our method in which we only classify an HVC detection based on its detected absorption in more than one species. Considering these sightlines but classifying them with no HVC detection or removing these sightlines from the counting would not change the results appreciably. We include these sightlines but do not count them as detections (see next in \S\ref{s-search}).

\subsection{Search for HVC absorption}\label{s-search}

\begin{figure*}
  \includegraphics[width=18 truecm]{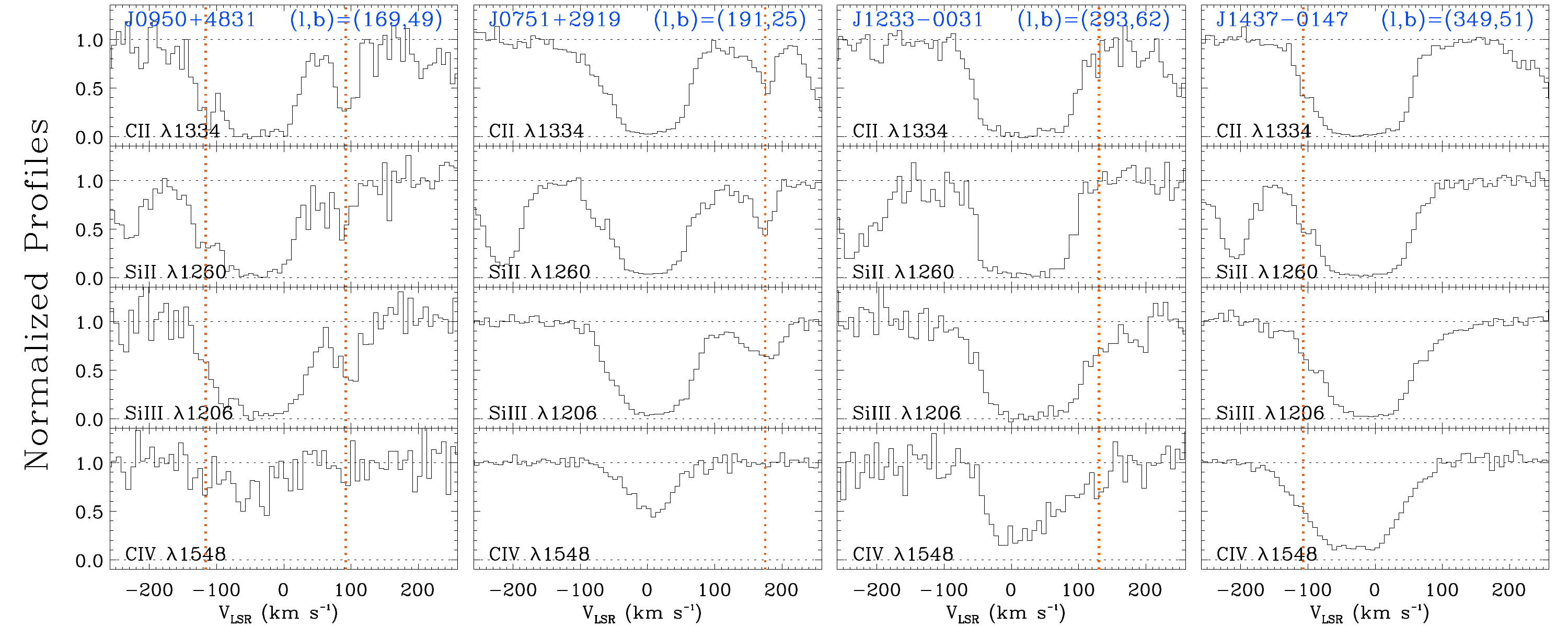}
  \caption{Examples from the TT sample showing the variety of HVCs and S/N. The dotted vertical lines show the HVC detections. The only sightline that passes through an \hi\ HVC complex (A) is J0950+4831, but only the negative component is associated with complex A along this direction. There is no \hi\ HVC complex near the other directions (however, while complex L is several degrees away from J1437--0147, it has similar LSR velocities as seen toward this sightline). Also note how the ionization conditions can change with, e.g., in some cases absorption only in singly and doubly ionized species (J0751+2919) and in other cases  only in doubly and triply ionized species (J1233--0031).  
  \label{f-ex}}
\end{figure*}

In order to search for the high-velocity interstellar absorption ($90 \le |v_{\rm LSR}|\le 400$ \km) in the TT and GTO samples, we first used the strongest lines of atomic, singly, and doubly ionized species available in the COS UV bandwidth: \cii\ $\lambda$1334 ($f \lambda = 171$), \siii\ $\lambda$1260 ($f \lambda = 1487$), and \siiii\ $\lambda$1206 ($ f \lambda = 1967$). The smaller $f\lambda$  for \cii\ is compensated by the fact that C is about 8 times more abundant than Si in a standard solar abundance pattern. These ions have been shown to be extremely powerful for finding low \hi\ column density HVCs, and more precisely HVCs with virtually any \hi\ column densities (see below). The use of multiple transitions is crucial when using high redshift AGN because the likelihood of unrelated QSO absorber contamination is not negligible. Our rule is that an HVC absorption is defined as such only if it is seen in absorption in at least two species or atomic transitions. The same rule was used by LH11 for their stellar sample. To confirm the detection or to help determine the velocity of the HVCs, we also used tracers of neutral gas (e.g., \oi), weaker transitions of \siii, and tracers of more highly ionized gas (e.g., \civ). The strong line of the \civ\ doublet has $f \lambda = 147$, comparable to the strength of \cii, but \civ\ is more suitable than \cii\ if the gas is highly ionized. In Fig~\ref{f-ex}, we show four sightlines from the TT sample, highlighting the diversity of UV-selected HVCs. The absorption profiles in this figure were normalized using Legendre polynomials with degrees $d\le 3$.

\begin{figure}
  \includegraphics[width=9 truecm]{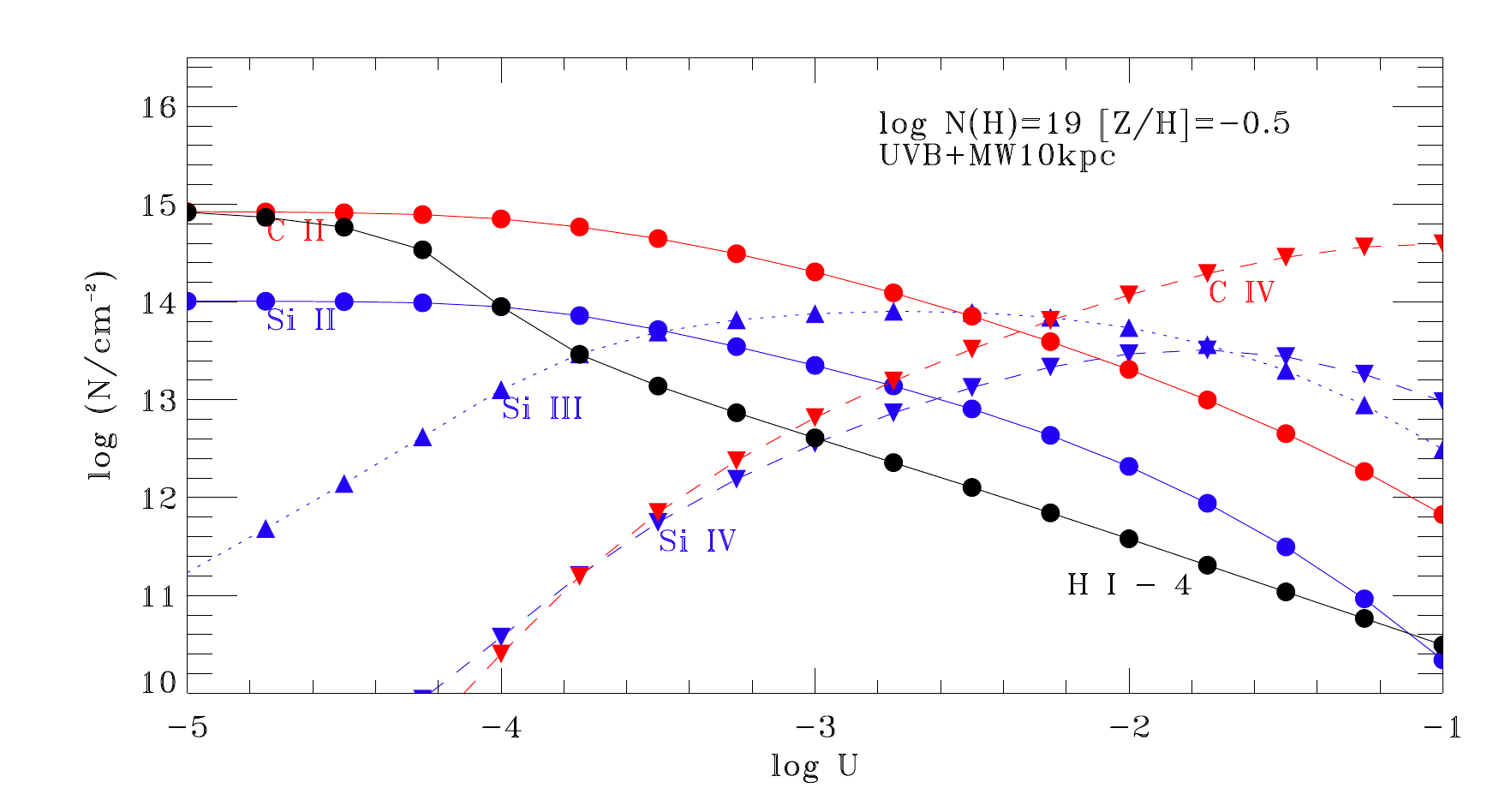}  
  \caption{Cloudy simulations predicting the column density of the observed ions against the ionization parameter $U = n_\gamma/n_{\rm H}$. The radiation field is from the combined, Milky Way plus UV background radiation field  at the position $R_G = 10$ kpc and $z = 5$ kpc. Note that in this Cloudy simulation $N({\rm H})$ is fixed and $N($\hi$)$ varies.   \label{f-mod}}
\end{figure}

Fig.~\ref{f-ex} shows that specific examples that the ionization conditions change fro sightline to sightline with in some cases only the detection of \cii,\ siii, and \siiii\ (no high ion), and in other cases only detection of \siiii\ and \civ\ (no low ion). To illustrate this change in the ionization conditions, we show in Fig.~\ref{f-mod} a Cloudy \citep{ferland98} photoionization model that predicts the column densities of \hi, \cii, \civ, \siii, \siiii, and \siiv\ as a function of ionization parameter ($U = n_\gamma/n_{\rm H}$, where $n_\gamma$ and $n_{\rm H}$ are the densities of ionizing photons and hydrogen). The radiation field is from the combined Milky Way and UV background radiation fields at a Galactocentric radius ($R_G$) of 10 kpc and a $z$-height of 5 kpc \citep[][]{fox05,fox10}.  These distances were chosen because they are close to the observed values  of the HVCs \citep[this paper, LH11,][]{thom06,thom08,wakker08,wakker07}.  We consider a total H column density of $10^{19}$ cm$^{-2}$ and a  metallicity of a 1/3 solar (see \S\ref{s-met}). It is evident from this figure that the combination of several ions of the same and different species is extremely powerful for finding HVCs; e.g., as $U$ increases, the amount of \cii\ diminishes and may become undetectable, but then \civ\ can be detected. Therefore the available UV diagnostics allow us to search sensitively for any types of HVCs, neutral, weakly, or highly ionized, and this is well illustrated in Fig.~\ref{f-ex}. Also note that the \hi\ column density decreases by a large amount from about $10^{19}$ to $10^{15}$ cm$^{-2}$ for $-5\le \log U\le  -1$. The total H column density is by definition always $10^{19}$ cm$^{-2}$ in this illustration.

The fourth column of Table~\ref{t-sample} summarizes the detection or non-detection of HVCs along the sight lines of the TT sample, including the velocity of detected HVC absorption. The velocities listed in this table were estimated from the apparent optical depth-weighted mean velocity  $\langle v\rangle = \int v \tau_a(v)dv /\int \tau_a(v)dv $. A colon in this column  highlights that the velocity is severely blended with the lower velocity gas. A velocity with the superscript $c$ indicates that the HVC was only detected in \siiii\ and \civ; there are only 5 such cases.

\subsection{Sensitivity}\label{s-sens}
Knowing the sensitivity of our sample is key for determining if the detection of the HVC is dependent on the S/N of the spectra and for comparing or combining our sample with other samples. Below we demonstrate that the TT sample is sensitive enough to detect HVC absorption at a level previously detected in other samples.  

\subsubsection{Sensitivity of the TT sample}
In Table~\ref{t-sample}, we report the 3$\sigma$ upper limits on the equivalent width ($W_{\rm lim}$) obtainable for each line of sight from the TT sample and display its distribution in Fig.~\ref{f-dist}. We estimated $W_{\rm lim}$ near the \siiii\ absorption, but the distribution of $W_{\rm lim}$ near \cii\ or \siii\ would be very similar (although the actual value of $W_{\rm lim}$ of \cii\ or \siii\ may be different from those listed in Table~\ref{t-sample} for a given sightline). To estimate $W_{\rm lim}$, we measured the equivalent width (and $1\sigma$ error) over the same velocity range ($\sim 80$ \km) that is usually observed when an HVC is detected.  The 3$\sigma$ upper limit on the equivalent width is defined as the 1$\sigma$ error times three. The mean and median of $W_{\rm lim}$ are 42 m\AA, and the dispersion around of the mean is $\sigma_d = 16$ m\AA. Assuming the absorption line lies on the linear part of the curve of the growth, $\langle W_{\rm lim}\rangle$ corresponds to $\log N($\cii$\, \lambda1334)\sim 13.3$, $\log N($\civ$\,  \lambda1548)\sim 13.0$, $\log N($\siii$\,  \lambda 1260)\sim 12.4$, $\log N($\siiii$\,  \lambda 1206)\sim 12.3$. 


\subsubsection{Comparison with previous samples of detected HVCs}
\begin{figure}
  \includegraphics[width=8.3 truecm]{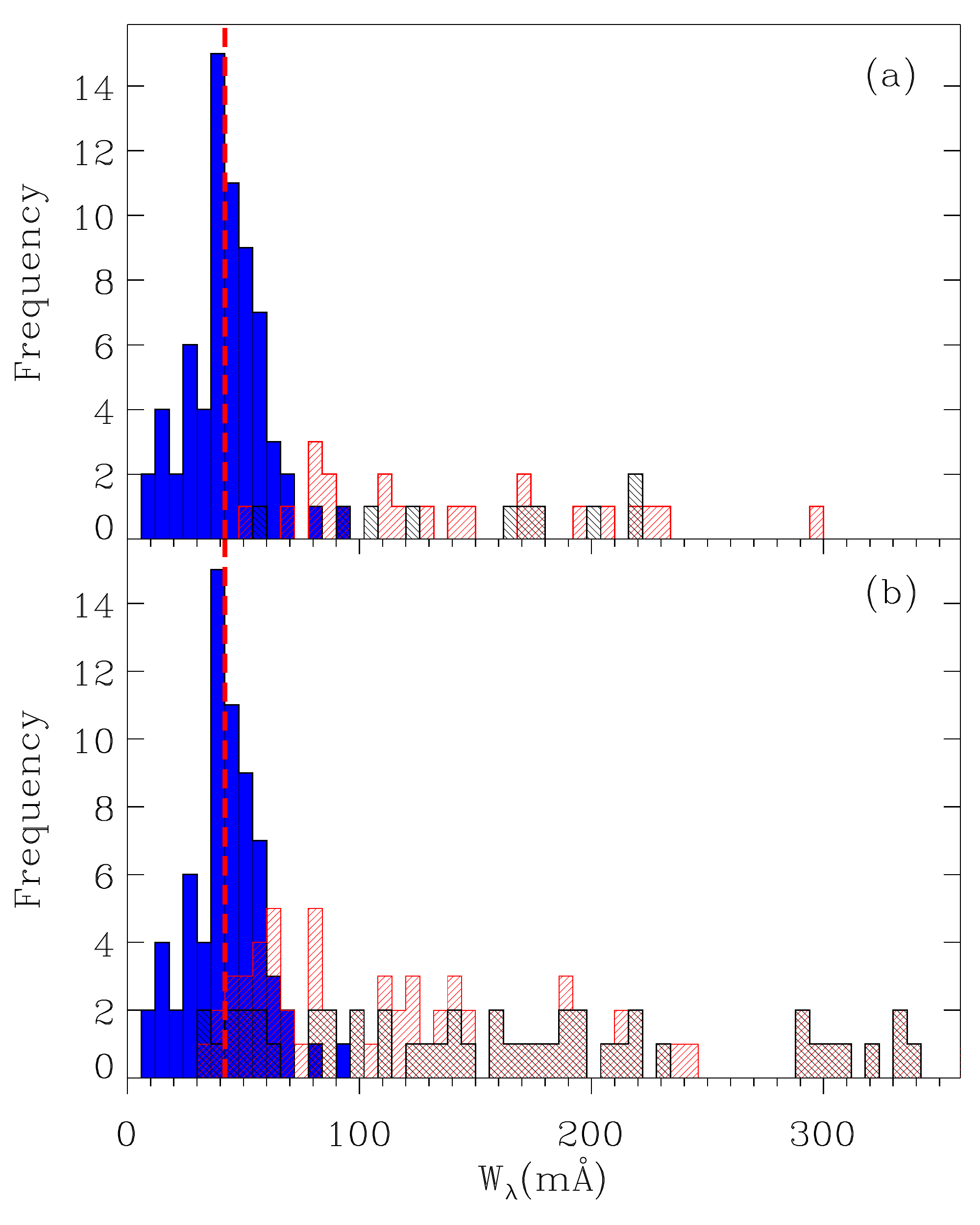}
  \caption{(a) Distribution of the HVC equivalent widths and equivalent width limits. The blue filled  histogram represents the $3\sigma$ equivalent width detection limit of the AGN TT sample, with the vertical red dashed line being the median (42 m\AA) of this distribution. The red hashed histogram is the \cii\ $\lambda$1334 equivalent width of the HVC detected in the TT AGN sample. The black hashed histogram is the \cii\ $\lambda$1334 equivalent width of the HVC detected in the LH11 stellar sample. (b) Same as (a), but the  red hashed histogram is the \siiii\ $\lambda$1206 equivalent width of the HVC detected in the CSG09 AGN sample and black hashed histogram shows the distribution of the strongest HVC component along each line of sight of the CSG09 AGN sample.  \label{f-dist}}
\end{figure}

In order to compare our results with others, we estimated the equivalent widths $W_\lambda$ of \cii\ for the HVC component when it is detected. We chose \cii\ because this ion is cleanly observed in the stellar spectra. For example, in stellar spectra, \siiii\ $\lambda$1206 cannot always be estimated owing to contamination by the star itself or because the blended interstellar and stellar Ly$\alpha$ absorption can severely depress the flux near \siiii. We did not attempt to separate severely blended HVC components and only measured $W_\lambda$ when the HVC absorption was not entirely blended with the lower velocity absorption. In Fig.~\ref{f-dist}, we show the distribution of $W_\lambda($\cii$)$ in the TT sample. The strength of most of the detected HVC absorption features is above  $\langle W_{\rm lim}\rangle+2\sigma_d$ and is strictly larger than  $\langle W_{\rm lim}\rangle$. 

Using the stellar sample of LH11, we similarly estimated $W_\lambda($\cii$)$ when a HVC is detected. Although the average S/N of the stellar spectra is higher than the S/N of the AGN spectra in the TT sample,  the equivalent widths of HVCs in the stellar sample are such that $W_\lambda \ga \langle W_{\rm lim}\rangle+2\sigma_d$ (see Fig.~\ref{f-dist}). This implies that the detection rates of the stellar and  AGN TT samples can be directly compared as we observe no HVC with  $W_\lambda < \langle W_{\rm lim}\rangle+2\sigma_d$.

Several AGN in CSG09 have spectra with S/N similar to the average S/N in the TT sample, but several of them have also higher S/N (see Figs.~1 and 2 in CSG09). In Fig.~\ref{f-dist}, we show the measured \siiii\ equivalent widths by CSG09 in the individual components of the HVCs of their sample. Following our guidelines, we removed any HVC that is identified only with \siiii. Except for 3 high-velocity components, all the detected  HVCs in CSG09 have also an equivalent width larger than $\langle W_{\rm lim}\rangle$. Another way to consider the CSG09 sample is for each sightline to retain only the strongest HVC absorption components. This is shown by the black histogram in Fig.~\ref{f-dist}. For either samples, we find that most of the detected HVC absorption features have  $W_\lambda> \langle W_{\rm lim}\rangle + 2\sigma$ and is larger than  $\langle W_{\rm lim}\rangle$.

The sensitivity of the GTO sample is typically similar to the best S/N in the TT sample (i.e., comparable to data in the GO program 11741). Yet, we estimate equivalent widths of the HVC absorption that are similar to those derived in the TT sample. Therefore despite a non-uniformity in the S/N, the TT, GTO, and CSG09 samples can be merged and can be compared to the LH11 stellar sample. Fig.~\ref{f-dist} shows that the detections are generally of high significance as otherwise weak absorption of  \siiii, \siii, \civ, \cii\ would have been detected more frequently in the sample with better sensitivity. This implies that low total H column HVCs are not important and that the present sample  is essentially complete.

\section{Covering factor and distance of the HVC\lowercase{s}}\label{s-cov}
\begin{figure}
  \includegraphics[width=8.8 truecm]{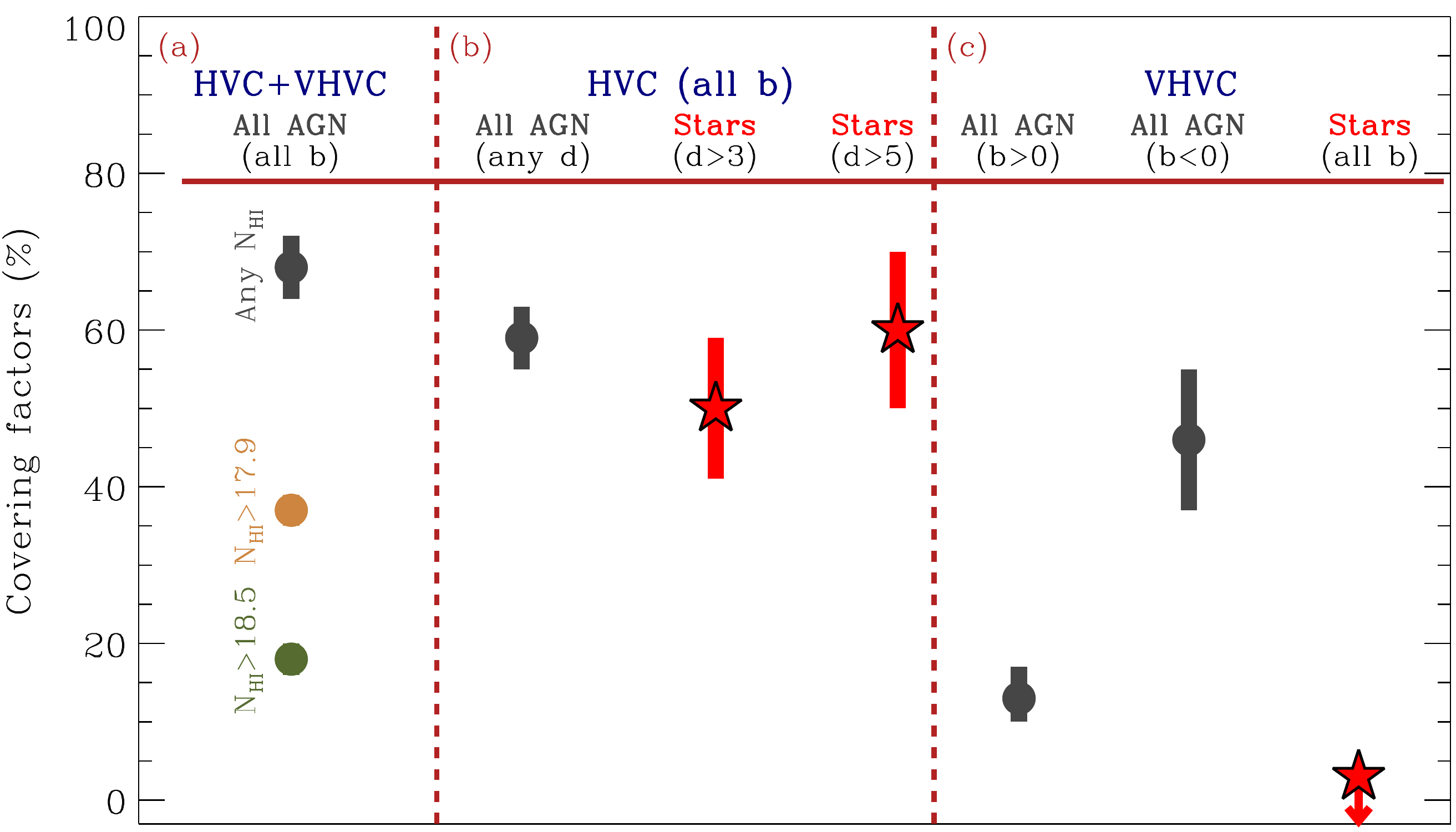}  
  \caption{Covering factors for the HVCs and VHVCs: (a) comparison between the \hi\ emission ($N($\hi$)$ limited) and UV absorption (any $N($\hi$)$, i.e., $N($\hi$)\la 10^{15}$  cm$^{-2}$ to $N($\hi$)\ga 10^{20}$  cm$^{-2}$) samples; (b) comparison the covering factors determined from the TT+GTO+CSG09 AGN and LH11 stellar (where two different heliocentric distances cutoff are considered at 3 and 5 kpc) samples; (c) VHVC covering factor from the TT+GTO+CSG09 AGN and LH11 stellar samples.  \label{f-sum}}
\end{figure}

The sky covering factor of HVCs is a crucial quantity, as it is required to statistically determine the distance of the UV absorption-line selected population of HVCs (LH11) and for estimating the total mass of the HVCs ($M \propto f_c d^2$, where $f_c$ is covering factor).  Following LH11, we define two classes of HVCs: the HVCs with $90 \le |v_{\rm LSR}| \la 170$ \km\ and the VHVCs with $|v_{\rm LSR}|>170$ \km. The reason for this separation in velocity is that HVCs were detected toward stars, but no VHVC absorption was found in the stellar spectra; the precise velocity cutoff is somewhat arbitrary (as is the lower bound at 90 \km, see \S\ref{s-vhvc}for more details). In Table~\ref{t-det}, we summarize the values of $f_c$ for different AGN samples and the LH11 stellar sample (see LH11 and \S\ref{s-dist}). In this table, $m$ is the sample size (i.e., the number of sightlines). We assess a 68\% confidence interval for each value of $f_c$ using the Wilson score interval for a binomial distribution. The main results can be visualized in Fig.~\ref{f-sum}, are listed in Table~\ref{t-det}, and can be summarized as follows: 
\begin{enumerate}
\item $f_c$ does not strongly depend on the sensitivity of the present data; 
\item Comparing the \hi\ surveys with ours, we conclude that 74\% of the HVC+VHVC directions have  $N($\hi$)< 3 \times 10^{18}$ cm$^{-2}$ and 46\% have $N($\hi$)<8 \times 10^{17}$ cm$^{-2}$ (panel (a) in Fig.~\ref{f-sum});
\item the HVC detection rates in the stellar and AGN samples overlap within $1 \sigma$, strengthening the LH11 conclusion that most of the HVCs are near the Milky Way disk, within about heliocentric distances 5--15 kpc (panel (b) in Fig.~\ref{f-sum});
\item  the HVC detection rate in the stellar sample drops with decreasing distances (panel (b) in Fig.~\ref{f-sum});
\item VHVCs are far more frequent at $b<0\degr$ than at $b>0\degr$, which is largely due to large Galactic sky covering of the Magellanic Stream at $b<0\degr$ (panel (c) in Fig.~\ref{f-sum}). 
\end{enumerate}
In the following sections, we explain and discuss these conclusions.

\begin{table}
\begin{minipage}{8.5cm}
\begin{center} 
\caption{Covering factors  for the HVCs and VHVCs \label{t-det}}
\begin{tabular}{lcccc}
  \hline
   	  & $m$   &  $f_c({\rm HVC})$		&   $f_c({\rm VHVC})$    	     &   $f_c({\rm HVC+VHVC})$	          \\
   	  &  &  $(\%)$		&    $(\%)$	    	     &    $(\%)$	          \\
 \hline
\multicolumn{5}{c}{TT AGN Sample with $W_{\rm lim} \ge 10$ m\AA\ } \\
 \hline
all $b$			&  67 & $    58 \pm 6	$ & $    6 \,^{+  4}_{-  2} $ & $   63 \pm 6 $ \\
$b>0\degr$    		&  60 & $    58 \pm 6	$ & $    3 \,^{+  3}_{-  2} $ & $   62 \pm 6 $ \\
 \hline
\multicolumn{5}{c}{TT AGN Sample with $10 \le W_{\rm lim} \le 42$ m\AA\ } \\
 \hline
All $b$			&  34 & $    59 \pm 9	$ & $    6 \,^{+ 5}_{-  3} $ & $   62 \pm 9 $ \\
$b>0\degr$    		&  31 & $    58 \pm 9	$ & $    7 \,^{+ 6}_{-  3} $ & $   61 \pm 9 $ \\
 \hline
\multicolumn{5}{c}{TT+GTO+CSG09 AGN sample } \\
 \hline
all $b$			&  133 & $    59 \pm 4	$ & $    20 \pm 4 $ & $   64 \pm 4 $ \\
$b>0\degr$    		&  105 & $    57 \pm 5  $ & $    13 \pm 3 $ & $   61 \pm 4 $ \\
$b<0\degr$    		&  28  & $    68 \pm 9  $ & $    46 \pm 10$ & $   75 \,^{+ 7}_{-  9}$ \\
 \hline
\multicolumn{5}{c}{LH11 stellar sample } \\
 \hline
all $b$, $d> 3$ kpc	&  28 & $    50 \pm 9	$ & $    0\,^{+  3}_{- 0} $ & $   50 \pm 9$ \\
all $b$, $d> 5$ kpc	&  20 & $    60 \pm 10	$ & $    0\,^{+  5}_{- 0} $ & $   60 \pm 10$ \\
 \hline
\end{tabular}
\end{center}
Note: $m$ is the size of the considered sample. Errors are 68.3\% level confidence interval estimated from the Wilson score. 
\end{minipage}
\end{table}

\subsection{HVC covering factor from the AGN samples}\label{s-all}

We first consider solely the TT sample because we are certain that no bias was introduced in selecting the targets that would favor or disfavor the detection of HVCs. In order to confirm that our HVC samples are essentially complete, we consider one sample with $W_{\rm lim} \ge 10$ m\AA\ (i.e., the entire TT sample) and a sample with $10 \le W_{\rm lim} \le 42$ m\AA\ (i.e, a sample with a higher sensitivity). The results are summarized in Table~\ref{t-det}. The entire and more sensitive TT samples show that $f_c$ is essentially independent of the equivalent width sensitivity of the present data (a sample with $W_{\rm lim} \ge 42$ m\AA\ gives $f_c =56\%$, similar to the $f_c$-values listed in Table~\ref{t-det}). As demonstrated in Fig.~\ref{f-dist} and discussed in \S\ref{s-sens}, there is no evidence for a large population of weak metal absorbers with a significant $f_c$. 

\begin{figure}
  \includegraphics[width=8.5 truecm]{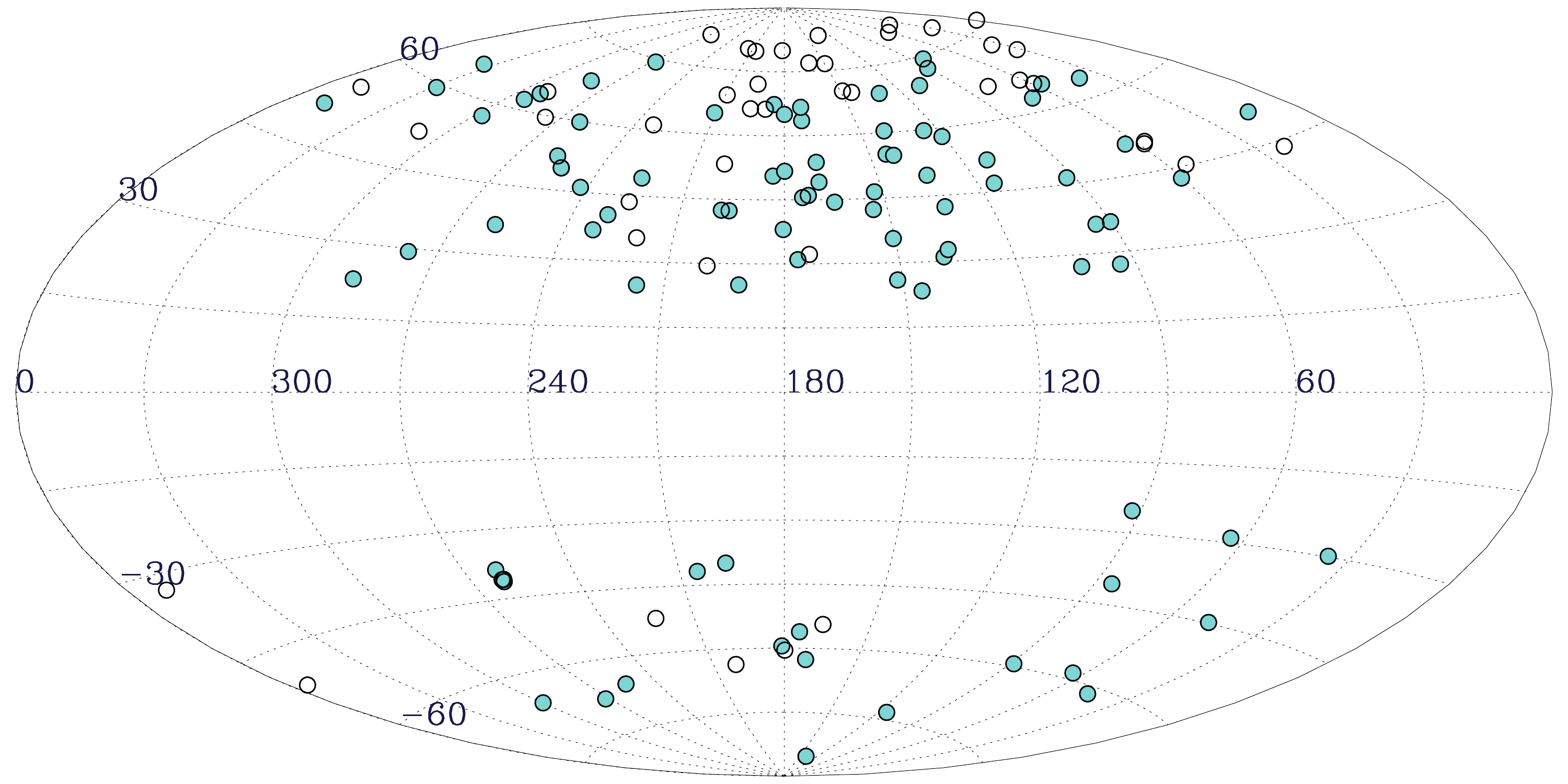}
  \caption{Aitoff projection of the combined (TT+GTO+CSG09) sample. A filled circle means at least one HVC with $|v_{\rm LSR}|>90$ \km\ is detected along the line of sight. An open circle means that no HVC is detected along the sightline.
  \label{f-map1}}
\end{figure}

Fig.~\ref{f-map} shows that there is no HVC at $b>75\degr$ in the TT sample, which is also confirmed with the combined sample (see Fig.~\ref{f-map1}). For $30\degr<b<75\degr$ the HVC detection rate is necessarily higher at $69\%$ compared with $58\%$ for the $b>0\degr$ sky. The absence of HVCs at $b>75\degr$ could be due in part to statistical fluctuations over  the small solid angle covered by these latitudes. If the HVCs follow a thickened disk-like distribution, a deficit of clouds at $b>75\degr$ would also be more likely. 

In Fig.~\ref{f-map1}, we show the distribution of sightlines and HVC detections for the combined sample that includes the TT, GTO, and CSG09 samples. This combined sample has much better coverage at $b<0\degr$ where it is larger than the TT sample by a factor 4. The observations are still heavily weighted toward the northern Galactic sky. In Table~\ref{t-det}, we summarize the detection rates of the combined sample. We again consider the HVC, VHVC, and HVC+VHVC categories in 3 sub-samples: the entire Galactic sky, the northern Galactic hemisphere ($b>0\degr$), and the southern Galactic hemisphere ($b<0\degr$). 

For $b>0\degr$, the hit rates are very similar for the HVC and HVC+VHVC between the TT and full samples. Within the errors the VHVC hit rate is also similar and small at $b>0\degr$. Comparing the entire sky yields similar conclusions, but again the sample is heavily weighted to $b>0\degr$ with $\sim 4$ times more AGN than at $b<0\degr$. Given this similarity, we will adopt the results from the combined sample for the remainder of this work. There is, however, a clear difference at $b<0\degr$: the covering factor of the VHVC is substantially larger, with a 46\% covering factor compared to 13\% at $b>0\degr$ (see Table~\ref{t-det} and Fig.~\ref{f-sum}). There is also a tentative increase in $f_c({\rm HVC})$ at $b<0\degr$ compared to that at $b>0\degr$, although the values overlap with the errors. As we discuss further in \S\ref{s-vhvc}, the southern Galactic sky is dominated by the Magellanic Stream \citep{putman98,nidever08}, and this mostly explains the considerable difference in $f_c({\rm VHVC})$ between the two Galactic hemispheres. We defer a discussion on the sky coverage of the VHVCs and implications for the origins and distances of the VHVCs to \S\ref{s-vhvc}.

The covering factors derived from our samples using \oi, \cii, \civ, \siii, \siiii, \siiv\, etc.,  are quite consistent with those determined for HVCs with \ovi\ absorption, which are in the range of 60\%--75\%  \citep{sembach03,fox06}. The fact that the covering factors are similar is not surprising in view of the multiphase nature of the HVCs, which often show absorption by  \oi, \cii, \ciii, \civ, \ovi\ at similar velocities in a given sightline \citep[e.g.,][]{ganguly05,collins05,fox04,fox06,zech08}. As warm HVCs move in the Galactic halo, some of the ``\ovi\ HVCs" are likely the collisionally ionized interface between the hot Galactic corona and the cooler ionized and neutral HVCs \citep[e.g.,][]{sembach03,collins05,kwak11}.   

We note that the difference in the hit rates between our analysis and CSG09 \citep[and][where they found $f_c({\rm HVC+VHVC}) = 80$--$90\%$ based on the search of \siiii\ HVC absorption in AGN spectra]{shull09} is due to three causes: i) there was a bias in the CSG09 sample where AGN initially targeted for HVC science were not removed, which artificially increased the covering factor; ii) while CGS09 confirmed in many cases the HVC absorption with ions other than \siiii, in several cases they rely solely on the \siiii\ absorption and {\it assumed}\ that the intergalactic contamination is small; and iii) as we have just demonstrated, the TT and combined samples are not uniformly distributed in both hemispheres, thus our sample weights the northern hemisphere more strongly than CSG09 (and the CSG09 sample is more balanced between the two Galactic hemispheres). A uniform measure of the entire Galactic sky covering factor of the HVC+VHVC would likely be somewhat higher than the 64\% derived here. In view of this imbalance, a straight average between the two Galactic hemispheres is more appropriate, yielding a covering factor for the entire sky of $f_c({\rm HVC+VHVC}) = 68\% \pm 4\%$. That is the value adopted in Fig.~\ref{f-sum} (panel (a)).

\subsection{Distance of the HVC population}\label{s-fcd}

Having set a robust HVC covering factor of $f_c({\rm HVC}) = 59 \% \pm 4\%$ toward the AGN sightlines, we can revisit the LH11 method to statistically determine the distance of the HVC population by comparing the HVC covering factors determined  toward sightlines with no distance limits (AGN) and with distance limits (stars). With their stellar sample, LH11 found $f_c({\rm HVC}) = 50 \% \pm 9\%\, (m=28)$. The $f_c$-values of the AGN and LH11 stellar samples are therefore consistent within the 68\% confidence interval (see also Table~\ref{t-det} and Fig.~\ref{f-sum}). The same ions and techniques were used to search for the HVC absorption in the stellar and AGN spectra, although finding HVC absorption in the stellar sample often relied only on \cii\ and \siii\ (\siiii\ was often too noisy and \civ\ could be blended with the stellar absorption). While it may be possible for more highly ionized HVCs to be missed in the stellar sample, this is unlikely to be important since a large majority of the HVCs would still show absorption in \cii\ and \siii\ as discussed in \S\ref{s-search}.

Although the results are consistent within a 68\% confidence interval, the difference between the stellar and AGN $f_c$ values is large enough that it could suggest a fraction of HVCs being at much greater distances if $f_c$ is not distance-dependent. However,  owing to interstellar drag on these clouds, HVCs should slow down as they fall onto the Milky Way disk  \citep{benjamin97}.  More recent hydrodynamical simulations have also followed the infall of HVCs through the Galaxy and predict that the HVCs become disrupted by their interaction with the corona \citep[e.g.,][]{bland07,peek08,heitsch09,kwak11,joung11}. These two arguments imply that the number of HVCs should decrease at smaller $d$ or $z$.  As  LH11 selected their stellar sample so that the stars are at $|z|\ga 3$ and $d>3$ kpc (only one star was at $z = -2.6$ kpc, but $d=8.7$ kpc), we searched MAST for stars at smaller distances $d>2 $ kpc (setting this limit to avoid probing only our local environment) and $|z|>1$ kpc \citep[and $|b|>15\degr$ to avoid probing HVCs linked to phenomena occurring in the Galactic disk, see, e.g.,][]{lehner11a} to study the dependence between $f_c$ and $d$. We found an additional 8 stars that satisfy these criteria with the appropriate coverage of at least \cii\ $\lambda$1334, \siii\ $\lambda$1260, \siiii\ $\lambda$1206. We   summarized the full (LH11 plus 8 stars) stellar sample in Table~\ref{t-star1}, where the stars are ordered in $z$-height from the Galactic plane. 

\begin{figure}
  \includegraphics[width=8.8 truecm]{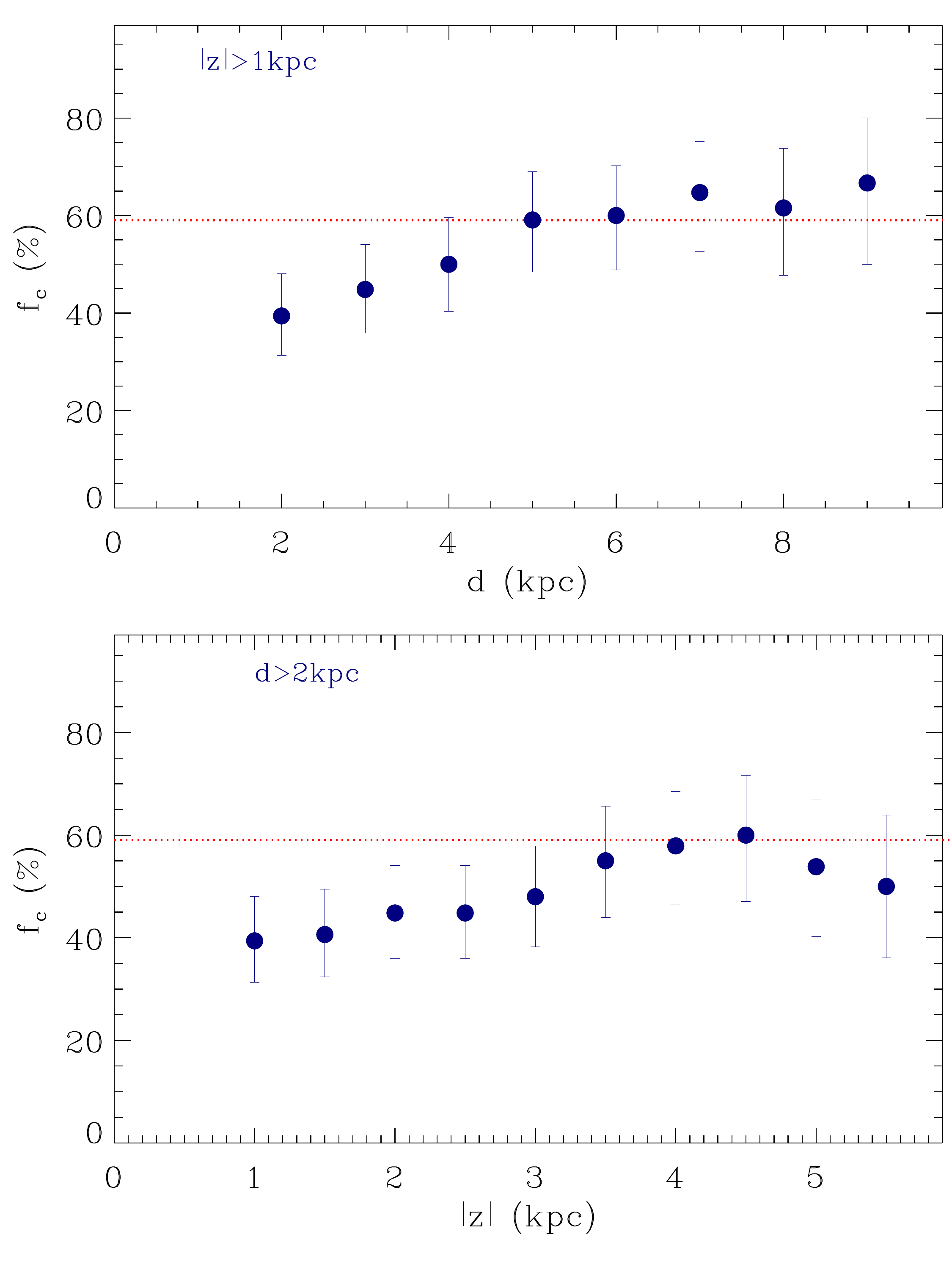}
  \caption{The HVC covering factors  toward the stars in the stellar sample summarized in Table~\ref{t-star1}, which includes the LH11 stellar sample plus 8 stars at $1<|z|<3$ kpc. {\it Top panel}: $fc$ vs. d. Each data point corresponds to the value of $f_c$ for stars at distances $>d_i$, where $d_i = 2, 3,...$ kpc. The horizontal dotted line is the HVC covering factor determined toward the AGN.  {\it Bottom panel}: same as top panel but  $fc$ is shown against $z$. As stars with smaller $d$ or $|z|$ are removed from the sample, $f_c$ converges on the $f_c$-value determined from the AGN sample. 
  \label{f-fcstar}}
\end{figure}

With this sample of stars, we estimate the values of $f_c$ as function of $d$ or $|z|$, i.e., we calculated $f_c$ for stars with $d> 2$ kpc (at any $|z|> 1$ kpc), $>3$ kpc, etc., and similarly with $|z|$ (and any $d>2$ kpc). The results are  summarized in Fig.~\ref{f-fcstar} where there is a distinct trend between $f_c$ and $d,z$. At $d\la 3$--4 kpc ($|z|\la 2$--3 kpc), $f_c$ is smaller and not consistent within the $68\%$ confidence interval with $f_c$ determined from the AGN sample, but as $d$ ($|z|$) increases,  $f_c$ increases, and plateaus at $d>4$ kpc ($|z|>3$ kpc) to the $f_c$ value determined from the AGN sample. For $d\ge 5$ kpc ($|z|>3$--4 kpc), $f_c =61\% \pm 10\%$, in  remarkable agreement with the result from the AGN sample.  Based on this result,  if we  set a cutoff at $z = 4$ kpc, we find: 
\\
-- $|z|\ge 4$ kpc, $f_c = 58\% \pm 11\%$ ($\langle d \rangle = 11.5 \pm 6.5$ kpc); \\
-- $|z|<4$ kpc, $f_c = 20\% \,^{+12\%}_{-8\%}$ ($\langle d \rangle = 4.1 \pm 1.8$ kpc).\\
A larger sample of stars would be needed to better characterize the infall of HVCs onto the Galactic disk, but our results strongly suggest that the population of HVCs decreases with decreasing $|z|$ (and $d$), and hence the HVCs may either be disrupted or decelerated to become intermediate-velocity clouds (IVCs, $50\le v_{\rm LSR} <90$ \km) or low-velocity clouds. A detail description of the (UV-selected) IVC population is beyond the scope of this work, but it would be extremely valuable to undertake to understand how HVCs may reach the Galactic disk. 

\begin{center}
\begin{table}
\begin{minipage}{8.2 truecm}
\caption{Stellar sample with stars at $d>2$ kpc, $|z|\ge1$ kpc \label{t-star1}}
\begin{tabular}{lccccc}
\hline
\hline
Name    &   $l$       &   $b$     & $d$  & $|z|$ & $v^{\rm HVC}_{\rm LSR}$   \\
        &   ($\degr$)&   ($\degr$)&  (kpc) & (kpc) &(\km) 	 \\
\hline
      HD195455$^a$   & $     61  $ & $ -27      $ & $ 2.2 $ & $   1.0 $ &	 $+94 $  	   \\		
      HD116852$^{a}$   & $    304  $ & $ -16      $ & $ 4.5 $ & $   1.3 $ &	 \nodata  		   \\	
      HD215733$^a$   & $     85  $ & $ -36      $ & $ 2.9 $ & $   1.7 $ &	 \nodata$^b$    	   \\		
      HD149881$^a$   & $     31  $ & $  +36      $ & $ 2.9 $ & $   1.7 $ &	 \nodata 		   \\	
         JL212$^a$   & $    303  $ & $ -61      $ & $ 2.2 $ & $   1.9 $ &	 \nodata  		   \\	
NGC6723-III60	 & $     0$ & $   -17	    $ & $ 8.7 $ & $   2.6 $ & $  -90  $   	   \\
      HD100340$^a$   & $    258  $ & $  61      $ & $ 3.0 $ & $   2.6 $ &	 \nodata  		   \\	
       HD18100$^a$   & $    217  $ & $ -62      $ & $ 3.1 $ & $   2.7 $ &	 \nodata  		   \\	
      HD119608$^a$   & $    320  $ & $  43      $ & $ 4.1 $ & $   2.8 $ &	 \nodata		   \\	
      HD121968   & $    334  $  & $  +55     $ & $ 3.8 $ & $   3.1 $ &	 \nodata	 	   \\
PG1511+367	 &   $  59   $  & $   +59	$ & $  3.8 $& $  3.2 $ &  \nodata 		  	   \\
NGC104-UIT14     &   $ 306   $  & $  -45	$ & $  4.5 $& $  3.2 $ &	\nodata 	  	   \\
HD233622	 &   $ 168   $  & $  +44	$ & $  4.7 $& $  3.3 $ &  \nodata 		  	   \\
EC10500-1358     &   $ 264   $  & $  +40	$ & $  5.2 $& $  3.3 $ & $+97    $ 	   \\
PG1704+222	 &   $  43   $  & $   +32	$ & $  6.9 $& $  3.6 $ &  \nodata		  	   \\
PG0855+294	 &   $ 196   $  & $ +39 	$ & $  6.5 $& $  4.1 $ & $+93,+107$       \\
PG2219+094	 &   $  73   $  & $   -40	$ & $  6.6 $& $  4.2 $ &  \nodata 		  	   \\
PG0832+675	 &   $ 148   $  & $  +35	$ & $  7.5 $& $  4.3 $ & $  -123 $        \\
PG0955+291	 &   $ 200   $  & $  +52	$ & $  5.5 $& $  4.3 $ &	\nodata 	       \\
NGC6341-326	 &   $  68   $  & $   +35	$ & $  8.2 $& $  4.7 $ & $  -94: $  	   \\
NGC6205-Barnard29&   $  59   $  & $  +41	$ & $  7.7 $& $  5.0 $ & $ -107 	  $    \\
NGC5904-ZNG1	 &   $  4	 $  & $  +47	$ & $  7.2 $& $  5.3 $ & $  -140,-120$    \\
PG1610+239	 &   $  41   $  & $  +45	$ & $  8.4 $& $  5.9 $ &	 \nodata	  	   \\
HS1914+7139	 &   $ 103   $  & $  +24	$ & $ 14.9 $& $  6.0 $ &  $-175,-118$    \\
PG0122+214	 &   $ 133   $  & $  -41	$ & $  9.6 $& $  6.2 $ & $ -160,-91   $   \\
PHL346	  	 &   $  41   $  & $  -58	$ & $  8.7 $& $  7.4 $ &	 \nodata	  	   \\
SB357            &   $ 301   $  & $  -81	$ & $  7.9 $& $  7.8 $ &	\nodata 	  	   \\
PG0914+001	 &   $ 232   $  & $  +32	$ & $ 16.0 $& $  8.4 $ & $+100,+170  $    \\
PG0009+036	 &   $ 105   $  & $  -58	$ & $ 10.8 $& $  9.1 $ &	 \nodata	  	   \\
vZ\,1128    	 &   $  43   $  & $  +79	$ & $ 10.2 $& $  10.0$ &	\nodata 		   \\
PG1708+142	 &   $  35   $  & $  +29	$ & $ 21.0 $& $  10.0$ &	 \nodata	  	   \\
PG1002+506	 &   $ 165   $  & $  +51	$ & $ 13.9 $& $  10.8$ & $  -102,+101$    \\
NGC5824-ZNG1     &   $ 333   $  & $  +22	$ & $ 32.0 $& $  12.0$ & $ -160: $  	   \\
PG1323-086       &   $ 317   $  & $  +53	$ & $ 15.8 $& $  12.6$ & $ -91  	  $    \\
\hline
\end{tabular}
Note: $a$: Stars not originally in the LH11 stellar sample. $b$: There is an IVC at $v_{\rm LSR} =-84$ \km\ \citealt{fitzpatrick97}.  A colon means the result is tentative in view of the low signal-to-noise of the data (removing these stars from the sample would not change our conclusions). 
\end{minipage}
\end{table}
\end{center}

The excellent agreement in the HVC detection rates between the stellar and AGN samples strengthens the conclusions reached by LH11: most of the HVCs are within 5--15 kpc of the sun. The drop in the population of HVCs with $|z|$ is consistent with models predicting their fall through the Galactic halo onto the disk.  All this implies that the HVCs are distant enough to have sufficient mass and close  enough that their infall rate balances the star formation rate in the disk of the Milky Way  (see LH11 and \S\ref{s-infall}).  In \S\ref{s-dist} we also show that several of the ionized HVCs are closely related to the large \hi\ complexes given their proximity on the celestial sphere and in three dimensional space. The (neutral and ionized) HVCs are therefore very likely a large part (if not all) of the source of gas required for continued star formation.

\subsection{HVC covering factors from \hi\ emission and UV absorption data}
Searching for high-velocity \hi\ emission with $|v_{\rm LSR}| \ge 90$ \km\ using the LAB survey toward the sightlines in the combined sample, we find $f_c \simeq 18\%$  for $\log N($\hi$) \ga 18.5 $. Given that more sightlines in the CSG09 and GTO samples go through large \hi\ HVC complexes (Magellanic Stream, complex C and complex A), it is not surprising that the \hi\ LAB HVC detection is higher in these samples ($f_c \sim 22\%$) than in the TT sample  ($f_c \sim 12\%$).  The detection rate from the combined sample is entirely consistent with that derived for the entire radio \hi\ sky survey at a similar sensitivity \citep{wakker91}. We emphasize this result as the techniques are different and the absorption-line method samples comparatively sparsely the Galactic sky. Our AGN sample and the absorption-line technique are well suited for finding HVCs with both high and low \hi\ column densities. 

Deeper radio \hi\ emission surveys reached a sensitivity of $\log N($\hi$) \simeq 17.9 $ \citep{murphy95,lockman02}. At this limit, 37\% of the sky is covered by \hi\ HVCs+VHVCs, which is still substantially smaller than $f_c$ in our UV absorption survey.  Extrapolating  to $\log N($\hi$) \simeq 17 $, \citet{lockman02} argued that the HVC+VHVC covering factor would be about 60\%, close to our findings ($f_c = 68\%$). However, they assume that the constant covering factor per decade of  $\log N($\hi$)$ observed between 19 and 17.9 dex applies down to 17 dex. While  direct estimates of $ N($\hi$)$ via \hi\ absorption is difficult, \citet{fox06} searched for \hi\ Lyman series absorption associated with \ovi\ and/or \ciii\ HVCs. Their results show that the \hi\ column densities vary tremendously: the lowest column UV-selected HVCs  have $\log N($\hi$) \la 14.7$--$15.0$ while the highest have $\log N($\hi$) \ga 20$; thus there is at least a 5 dex range in $N($\hi$)$. The detection of \ovi\ or \ciii\ or any other metal line species in low \hi\ column density HVCs still implies $N({\rm H})>10^{17}$--$10^{18} (Z/Z_{{\odot}})^{-1}$ cm$^{-2}$. Thus, even though the \hi\ column density in these HVCs may be small, the total hydrogen column densities can still be substantial. 

Comparing the \hi\ surveys with ours, the relative covering factors imply about 74\% of the HVC+VHVC directions have  $N($\hi$)< 3 \times 10^{18}$ cm$^{-2}$ and 46\% have $N($\hi$)<8 \times 10^{17}$ cm$^{-2}$. We emphasize again the population of HVCs with extremely low total H column density is unimportant (see  \S\ref{s-sens}).

\begin{figure*}
  \includegraphics[width=16 truecm]{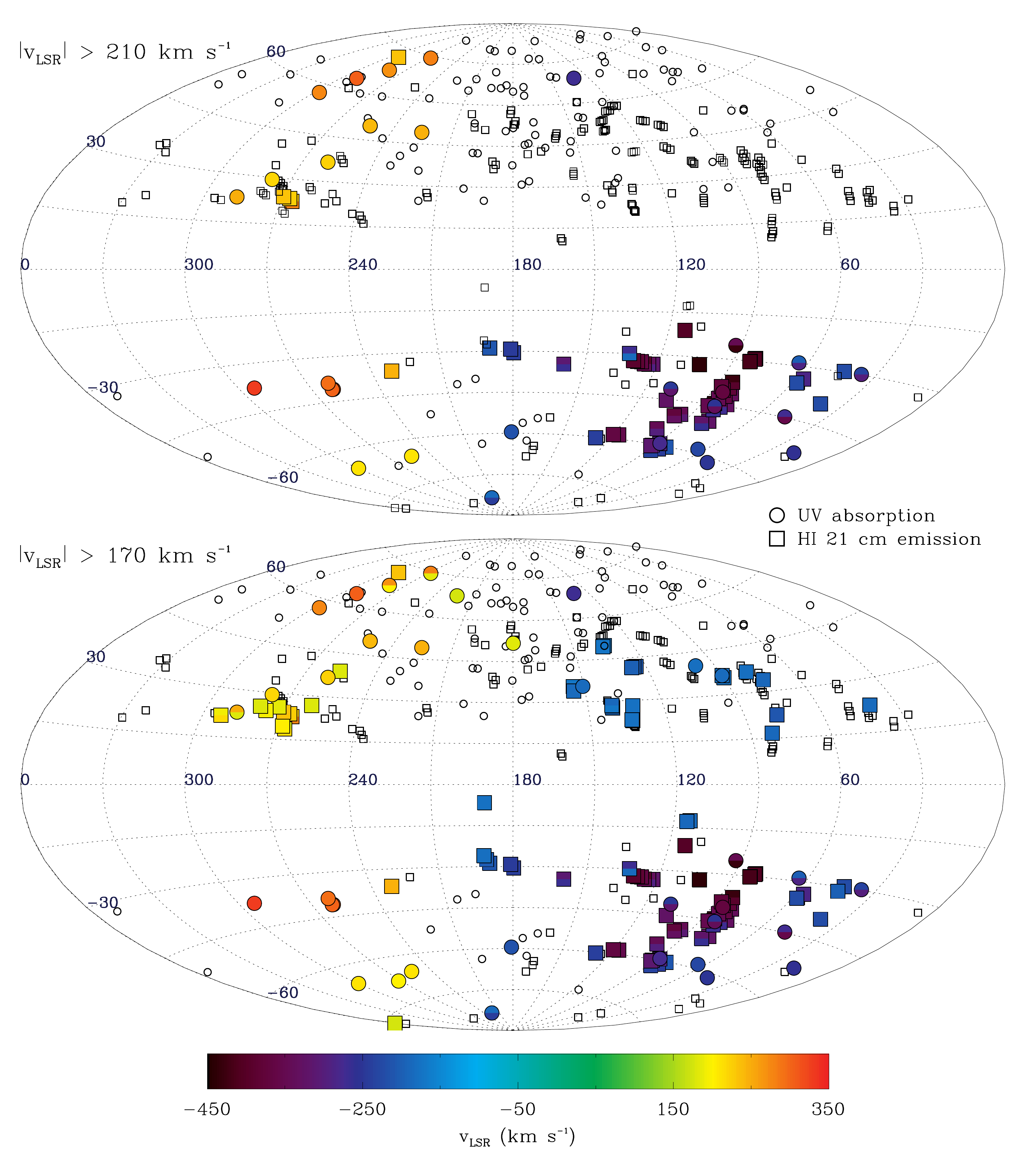}
  \caption{Aitoff projection maps showing the sightlines where a VHVC is detected toward an AGN (filled {\it circles}\ from UV absorption TT+GTO+CSG09 data; filled {\it square}\ from the \hi\ 21-cm emission data from Lockman et al. 2002). On the {\it top} panel, detections are VHVCs with $|v_{\rm LSR}|> 210$ \km, while on the {\it bottom}\ panel, detections are VHVCs with $|v_{\rm LSR}|> 170$ \km. The region defined by $b<0\degr$ and $240\degr<b<360\degr$ cannot be observed from Green Bank,  explaining the lack of \hi\ emission data in this region. All the open symbols indicate that no VHVC absorption or emission with  $|v_{\rm LSR}|> 210$ \km\ ({\it top}) or  $|v_{\rm LSR}|> 170$ \km\ ({\it bottom}) is found along the sightline.  
  \label{f-map2}}
\end{figure*}

\subsection{Covering factor and distribution of the VHVCs}\label{s-vhvc}
LH11 did not find any HVCs with $|v_{\rm LSR}| \ga 170$ \km, the so-called VHVCs, in absorption in the stellar spectra. They concluded that the VHVCs must be at distances $d>10$--$20$ kpc. Our analysis of the covering factors of VHVCs using the much larger sample of AGN demonstrates that at $b>0\degr$, the covering factor is quite small, $f_c ({\rm VHVC})\simeq 13\%$(see Table~\ref{t-det} and Fig.~\ref{f-sum}). In fact the TT sample nearly completely missed the VHVCs. The LH11 stellar sample and TT AGN sample are distributed similarly across the Galactic sky (see Fig.~\ref{f-map}), but the LH11 sample is about 2.5 times smaller. It may not be therefore surprising that no VHVC absorption was observed in the spectrum of a star ($f_c({\rm VHVC}) \la 13\%$ for a $95\%$ confidence interval for the stellar sample). Some of the stars cover the $b<0\degr$ hemisphere where the $f_c({\rm VHVC})$ is much larger, but, as we show below, most of the VHVC absorption is associated with the Magellanic Stream. The absence of Magellanic Stream absorption toward stars confirms that the Stream is indeed far away.  Using our large sample of AGN, we can learn more about the origins of these clouds as we now demonstrate.

\begin{table*}
 \centering
 \begin{minipage}{14.2cm}
  \caption{Heliocentric distance ($d_{\rm HVC}$), height ($|z_{\rm HVC}| = d_{\rm HVC}|\sin b|$), and Galactocentric distance$^a$ ($R_G$) of the HVCs and association with \hi\ HVC complex \label{t-dist}}
  \begin{tabular}{lcccccclc}
  \hline
{Name} & {$l$}& {$b$} & $d_{\rm HVC}$ & $|z_{\rm HVC}|$ & $R_G$  & $v^{\rm HVC}_{\rm LSR}$ & Associated \hi\ & Stellar Distance\\
 & ($\degr$)& ($\degr$) &  (kpc)& (kpc)& (kpc) & (\km)  & HVC Complex  & References\\
\hline
NGC6723-III60	 &     0      &    $   -17     $    &	   $<8.7$    &       $ <    2.6  $  &	   $< 0.2    $&   $  -90		 $	 &    None		 &  (1)   \\  
NGC5904-ZNG1	 &     4      &    $   +47     $    &	   $<7.5$    &       $  <   5.5  $  &	   $< 3.4    $&     $  -140	 $	 &    Complex L?	 &  (1)    \\  
NGC5904-ZNG1	 &     4      &    $   +47     $    &	   $<7.5$    &       $  <   5.5  $  &	   $< 3.4    $&     $  -120	 $	 &    Complex L?	 &  (1)    \\  
M13-Barnard29	 &     59     &    $	+41    $    &      $<7.1$    &        $ < 4.7	$    &     $< 7.4    $&    $  -121		 $	 &    Complex C 	 &  (1)   \\
HD195455$^b$     &     61     &     $ -27        $  &      $<2.2$    &       $  <   1.0  $   &	   $< 7.7$	      &    $+94 $  	                 & Complex gp?   &   (2)  \\		
HS1914+7139	 &    103     &     $  +24	$   &	 $<14.9$     &       $ < 6.0	 $  &	   $<17.6    $&    $-118		  $	  &    Outer Arm/Complex C? &  (3)   \\  
HS1914+7139	 &    103     &     $  +24	$   &	 $<14.9$     &       $ < 6.0	 $  &	   $<17.6    $&    $ -175		  $	  &    Complex C	  &  (3)    \\  
PG0122+214	 &    133     &     $  -41	$   &	  $<9.6$     &       $ < 6.2	 $  &	   $<14.5    $&    $ -91		  $	  &    Cohen Stream	  &  (3)   \\  
PG0122+214	 &    133     &     $  -41	$   &	 $<9.6$	     &       $ < 6.2	 $  &	   $<14.5    $&    $ -160		  $	  &	WW503		  &  (3)   \\  
PG0832+675 	 &    148     &     $  +35	$   &	  $\ge 8.1$  &       $ \ge 4.7	 $  &	   $\ge14.6  $&     $  -145		    $	    &	  Complex A	    &  (4)  \\  
PG0832+675 	 &    148     &     $  +35	$   &	 $< 8.1$     &       $ < 4.7	 $  &	   $<14.6    $&    $  -123	       $       &     Ionized complex A?&  (4) \\  
PG1002+506	 &    165     &     $  +51	$   &	  $<13.9$    &       $ < 10.8	 $  &	   $<17.1    $&    $  -102	       $       &    Complex M	       &  (5)  \\  
PG1002+506	 &    165     &     $  +51	$   &	  $<13.9$    &       $ < 10.8	 $  &	   $<17.1    $&    $  +101	       $       &    None	       &  (5)  \\  
PG0855+294	 &    196     &     $ +39	$   &	 $< 6.5	$    &       $ < 4.1	 $  &	   $<13.4    $&    $+93 	$      &    Cloud WW113        &  (6)	\\
PG0855+294	 &    196     &     $ +39	$   &	 $< 6.5	$    &       $ < 4.1	 $  &	   $<13.4    $&    $+107		$      &    Cloud WW113        &  (6)	\\
PG0914+001       &    232     &     $  +32 	 $   &   $< 16.0$    &        $< 8.4	 $   &     $<20.0    $&    $ +100		  $    &    Complex WB         &  (6)	\\  
PG0914+001       &    232     &     $  +32 	 $   &   $< 16.0$    &        $< 8.4	 $   &     $<20.0    $&    $ +170		  $    &    Complex WA         &  (6)	\\  
EC10500-1358     &    264     &     $  +40 	 $   &   $<  5.2$    &        $ <3.3	 $   &     $< 9.8    $&   $  +97		 $    &    Cloud WW95	      &  (2)  \\  
NGC104-UIT14     &    306     &     $  -45 	 $   &     $>4.5$    &        $ > 3.2	 $   &     $> 7.1    $&    \nodata$^c$  	      &    Complex WE	      &  (1)  \\  
PG1323-086       &    317     &     $  +53 	 $   &   $< 15.8$    &        $ <12.6	 $   &     $< 6.7    $&    $  -91		  $    &    None	       &  (7)	 \\  
\hline
\end{tabular}
($a$): Galactocentric distance of the HVCs, where $R_G = (R^2_{\odot} + (d_{\rm HVC}\, \cos b)^2  - 2 d_{\rm HVC} \, R_{\odot}\,  \cos l \, \cos b)^{0.5}$, assuming the distance from the Sun to the Galactic center $R_{\odot} = 8.5$ kpc. ($b$): Not in the LH11 sample. ($c$): No absorption is detected, but strong \hi\ emission is observed along this line of sight. 
References for the heliocentric distances of stars that were used to estimate the distance limits of the HVCs: (1) \citealt{harris96,harris10}; (2) \citealt{rolleston97}; (3) \citealt{ramspeck01}, (4) \citealt{wakker01}, (5) \citealt{ringwald98}, (6) \citealt{rolleston99}, (7)  \citealt{moehler98}. 
\end{minipage}
\end{table*}

\begin{figure*}
  \includegraphics[width=18 truecm]{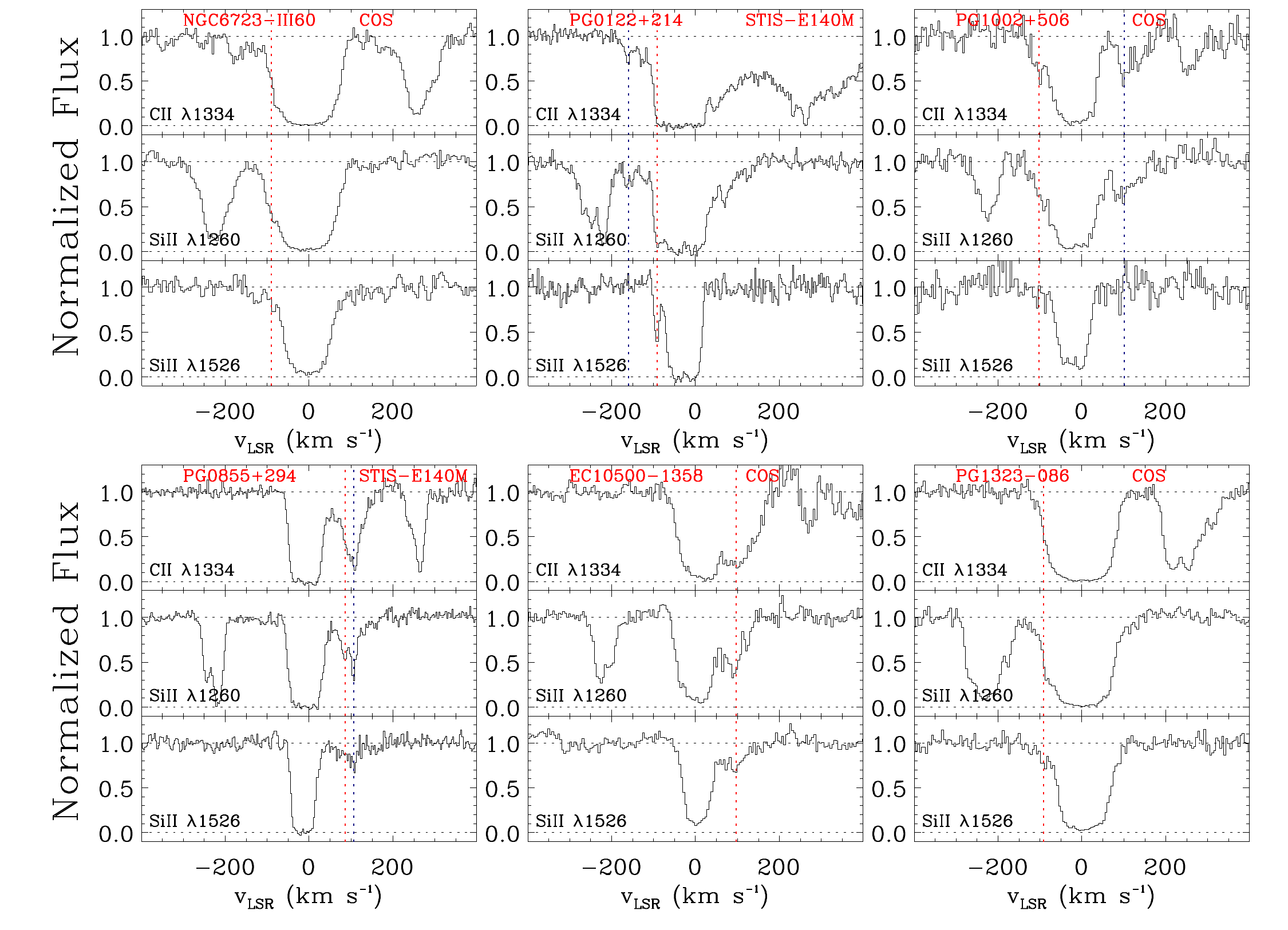}
  \caption{Normalized profiles  of interstellar \cii\ and \siii\ absorption seen in the spectra of several stars. The HVC components are marked by the vertical dotted lines. Other features are either stellar or interstellar.  The spectra were normalized within $\pm 600$ \km\ from the absorption lines using Legendre polynomials with a degree $d_p$ \citep[see, e.g.,][]{lehner11a}. For many stars, a fit with a low degree, $d_p=1$--3, was used, but some stars (e.g., PG0855+294, EC10500-1358) have complicated stellar profiles, especially near \cii, requiring higher values of $d_p>6$. There is a broad stellar line next the interstellar \cii\ feature at positive velocities in  the spectrum of EC10500-1358 that is only approximately modeled (a feature that is similar to that seen in PG0122+214 that we did not attempt to remove in this case as no HVC is seen at positive velocities in this direction). 
  \label{f-spec}}
\end{figure*}

 In Fig.~\ref{f-map2}, we show the sky distribution of VHVC absorption from our combined sample as well as \hi\ 21-cm emission from the  NRAO 140-ft ($\log N($\hi$)\ge 17.9$) observations of \citet{lockman02}. We do not show the lower velocity HVCs in order to highlight solely the VHVCs. In this figure, open symbols are used if no VHVC is seen in absorption or emission. In the bottom panel of this figure, the VHVCs with $|v_{\rm LSR}| > 170$ \km\ are shown, while the top shows only the VHVCs with $|v_{\rm LSR}| > 210$ \km. The overall sky coverage is excellent with only a lack of \hi\ data in the  $b<0\degr$ and $240\degr<l<360\degr$ quadrant because this region of the sky cannot be observed from Green Bank. The velocity distribution seen in emission and absorption is strikingly similar, which reinforces that the VHVCs seen in emission and absorption must be similar structures with different ionization levels (see \S\ref{s-dist}). It is therefore not surprising that there is a remarkable correlation between these maps and all sky \hi\ 21-cm HVC map  \citep[e.g.,][]{wakker03}, with many of the VHVCs superimposed onto or near large known \hi\ HVC complexes; the Magellanic Stream at similar velocities, complexes C and A, the anti-center HVCs at higher velocities. \citet{wakker01} and \citet{lockman02} attributed the gas moving at velocities with $-210 \la v_{\rm LSR} \la -170$ \km\ in the $b>0\degr$ and $0\degr<l<180\degr$ quadrant primarily to complex C. In the top panel of Fig.~\ref{f-map2}, where we only show now  the VHVCs with $|v_{\rm LSR}| > 210$ \km,  the VHVCs affiliated with complex C have almost completely disappeared  (as well as some of the anti-center complex). The VHVCs that remain present correlate extremely well with the \hi\ map of the Magellanic stream and its leading arm \citep{wakker03,nidever08}, except that the most sensitive \hi\ emission and UV absorption data suggest it could be wider (i.e., the Stream has a large ionized envelope) and could reach higher positive galactic latitudes than previously thought. Some of the VHVCs near $l=120\degr$--$130\degr$ and $b = -20\degr$ to $-30\degr$ are likely associated with the M31 and M33 galaxies \citep{thilker04,westmeier05,putman09}, but their solid angle is extremely small. We therefore conclude that most of the VHVCs with $|v_{\rm LSR}| > 210$ \km\ are dominated by the Magellanic Stream at $b<0\degr$ and its leading arm at $b>0\degr$. Only a very few of the VHVCs, such as those associated with M31 and M33, may not be part of the Stream. Independently, \citet{putman11} showed recently that many compact VHVCs have a head-tail structure (a signature of interaction with a hot diffuse medium) and can be associated with the Magellanic Stream. From our discussion, these small head-tail \hi\ VHVCs must have larger envelopes of photoionized gas and collisionally ionized gas. The VHVCs are therefore not randomly distributed on the Galactic sky and appear mostly to be associated to known large structures. However, while our AGN sample is large, the Galactic sky can only probed sparsely along pencil-beam sightlines with the absorption technique. It is therefore quite plausible that we might miss a population of compact HVCs with very small angular scales at very large distances (or that they may be confused with the more nearby VHVCs) as those possibly found with the ALFALFA survey \citep{giovanelli10}.  The compact HVCs are therefore the only remaining candidates for dark matter mini-halos in the HVC population.

\section{Association of the neutral and ionized HVC\lowercase{s} in 3D space}\label{s-dist}

Determining the distance of the HVCs has been critical for associating HVCs with flows occurring near the Milky Way rather than in the IGM of the Local Group and for quantifying their basic physical properties. The distances to many \hi\ HVCs have been determined using stars in direction of the \hi\ contours with $N($\hi$)\ge 3\times 10^{18}$ cm$^{-2}$. The high-velocity gas seen in absorption toward AGN outside the \hi\ contours has previously been associated with the extended HVCs seen in 21-cm emission based on their angular proximity and similarity in velocities \citep[e.g.,][]{fox04,collins05}. With the LH11 stellar sample, we can go a step further and directly compare the distances of \hi\ HVCs and HVCs seen in UV absorption. We emphasize that for most of the LH11 stellar sightlines, no \hi\ 21-cm emission (at the LAB survey sensitivity) is observed.

\begin{figure}
  \includegraphics[width=8  truecm]{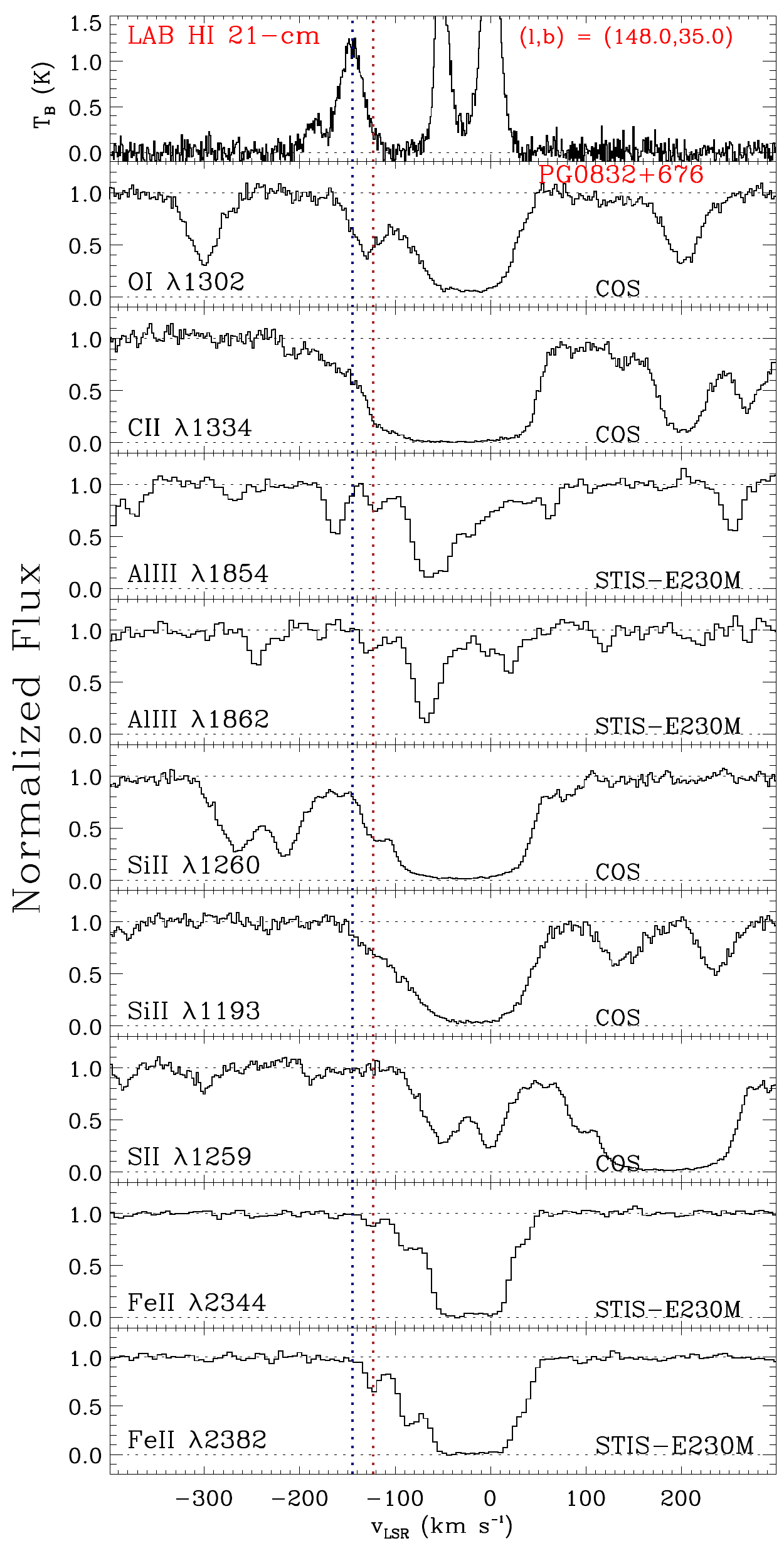}
  \caption{Normalized absorption profiles  of various metal-lines and LAB H\,I emission line profile ({\it top panel}) toward PG0832+675. The HVC components are marked by the vertical dotted lines. The main \hi\ emission is observed at $-145$ \km (note that the most negative emission component is not observed along this sightline using a $1\arcmin$ beam, Wakker et al. 2003) and is part of complex A. The main interstellar absorption is seen at $-123$ \km, although some absorption is also seen at $-145$ \km. The stellar spectrum is complicated with many stellar feature (the stellar velocity is $-70$ \km\ where absorption  is seen in \aliii\ as well as  \siiv\ and \civ). We argue that PG0832+675 must be at the front edge of complex A and the  $-123$ \km\ component could be part of its ionized envelope.
 \label{f-pg0832}}
\end{figure}

To search for \hi\ HVCs associated with the UV-selected HVCs, we used the \hi\ maps in \citet{wakker03} and \citet{wakker01} as well as information provided by B.P.~Wakker (private communication, 2011). In Appendix~\ref{s-app}, we summarize in detail the relation between the HVCs seen in UV absorption for each sightline with the \hi\ complexes, while in Table~\ref{t-dist}, we summarize the HVC distance constraints from the LH11 stellar sample for which the detection is unambiguously a high-velocity interstellar feature and the S/N was high enough for their reliable detection.\footnote{Specifically the following 4 stars were not considered: NGC6341-326  and NGC5824-ZNG1 owing to the low S/N in their COS spectra, and PG1243+275 and PG0934+145 for stellar contamination as noted by LH11. For the latter star,  there is a possibility that both stellar and interstellar high-velocity components are present at the same velocity.} In this table, we also added the additional star at smaller $d$ with a HVC detection (see \S\ref{s-fcd} and Table~\ref{t-star1}).  The heliocentric distance, $z$-height above and below the Galactic disk, Galactocentric distance, velocity, and  possible association with \hi\ HVC complexes are listed in Table~\ref{t-dist}. We show the normalized interstellar absorption spectra  in Figs.~\ref{f-spec}, \ref{f-pg0832}, \ref{f-spechd} for those stars whose STIS or COS spectra have not yet appeared in the literature (see Appendix~\ref{s-app}). We refer the reader to LH11 for more information about the data acquisition and reduction of the stellar sample. We emphasize that many of these HVCs are also detected in other transitions and/or species (e.g., \oi, \alii, \siiii, \civ, see figures in Lehner \& Howk 2010, LH11, Zech et al. 2008, and Fig.~\ref{f-pg0832}).

\begin{figure}
  \includegraphics[width=8 truecm]{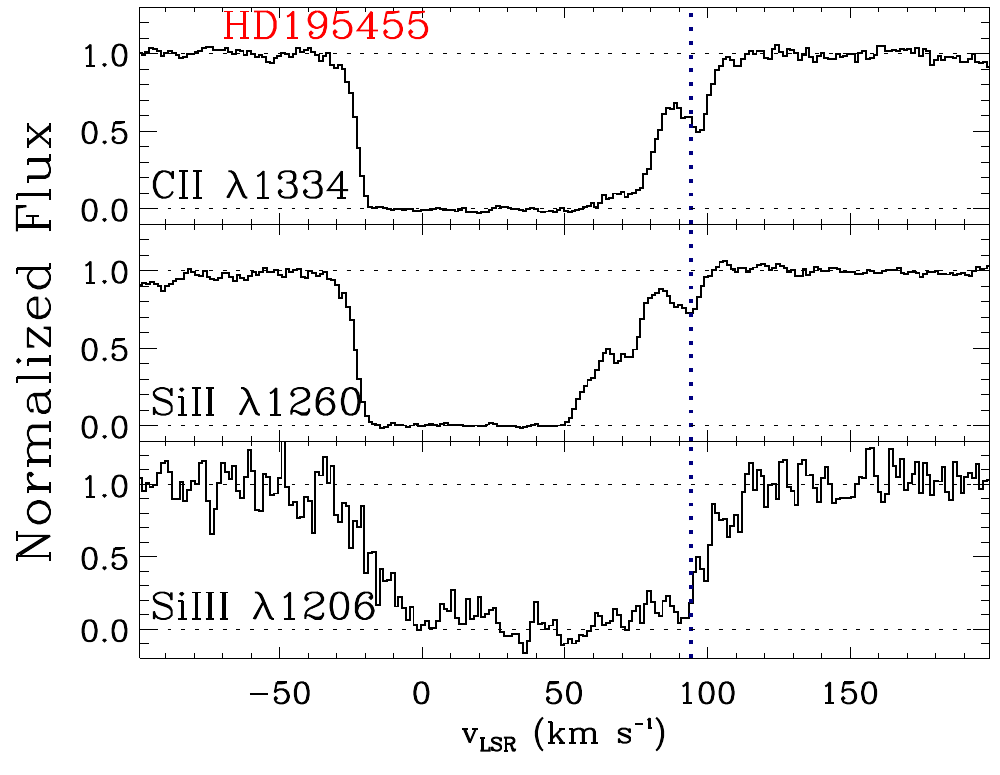}
  \caption{Normalized profiles  of interstellar \cii, \siii, \siiii\ absorption seen in the STIS E140H spectrum of HD195455, which was not originally in the LH11 sample. The component at $+70$ \km\  is consistent with the velocity of the IVC complex gp. The $+94$ \km\ component might be a higher velocity component of this complex. 
  \label{f-spechd}}
\end{figure}

From our detailed discussion in Appendix~\ref{s-app} and the results summarized in Table~\ref{t-dist}, we find that the distances to many of the predominantly ionized HVCs seen in the stellar spectra are consistent with the distances found for the \hi\ HVCs nearby on the sky at similar LSR velocities. This implies that many of the HVCs seen in absorption are the extended diffuse envelopes of the denser clouds seen in \hi\ emission. In view of the much larger covering factor of the predominantly ionized HVCs and the proximity of the sightlines to many \hi\ complexes, all the \hi\ HVCs are likely to have extended ionized envelopes. However, in both the stellar and AGN sightlines there are also ionized HVCs where no \hi\ complex is observed within $20\degr$--$30\degr$ of the sightline, implying that the ionized high-velocity gas is not only found near \hi\ complexes but also in regions devoid of \hi\ 21-cm emission (at least for  a \hi\ sensitivity of about $> 3\times 10^{18}$ cm$^{-2}$). 

\section{Implications}\label{s-imp}
\subsection{Origins of the gas flows in Milky Way halo}\label{s-met}

The results discussed in \S\ref{s-fcd} and \S\ref{s-dist} have demonstrated that:
\begin{enumerate}
\item most of the HVCs are within about heliocentric distances 5--15 kpc,
\item the population of HVCs drops with decreasing distances or $z$-height, and 
\item several HVCs seen in UV absorption are the extended ionized halos of their neutral counterparts.
\end{enumerate}
The  first item implies that  HVCs represent flows occurring in the Milky Way halo within about one Galactic radius from the Galactic center. The second item is consistent with infall of HVCs onto the Milky Way disk.  With the third item, it is not a big leap to conclude that the metallicities UV-selected HVCs are similar to those of their \hi\ counterparts. This is critical, as the metallicity can be difficult to estimate for the ionized HVCs, often requiring large ionization corrections \citep[although see][]{zech08}. Past and current estimates of the metallicities of HVCs find a wide range of values from about a few percent solar to super-solar metallicity \citep[e.g.,][B.P. Wakker, 2011, private communication]{wakker01,tripp03,collins03,fox04,zech08,shull11,yao11,tripp12}, with a rough average around  0.2--0.4 $Z_{\odot}$. As for their \hi\ counterparts, several ionized HVC complexes must therefore have an extragalactic origin, i.e., not all are recycled gas recently ejected from the Milky Way \citep[although we note that if some HVCs were ejected from regions situated at Galactocentric  radii beyond 10--15 kpc, a  metallicity of $0.4$ solar would not be unjustified based on the Galactic abundance gradient, e.g.,][]{chiappini01}. 

While some of these HVCs may be the result of cold stream accretion wherein  metal-poor gas flows onto galaxies along dense intergalactic filaments \citep{keres09,putman06} or remnants clouds from the Galaxy formation \citep{oort70,peek08}, it is also possible that much of this gas may come from the interactions with nearby dwarf galaxies and their outflows. Such a scenario may be seen farther away with the Magellanic Stream and Large Magellanic Cloud outflows. The masses of the Stream and outflows from the LMC are quite substantial \citep{putman06,nidever08,lehner09}. Being so far away, it is unclear if these HVCs will be able to reach the Galactic disk, instead they may mostly enrich the hot Galactic corona  \citep{fox10,putman11}. However, it is not inconceivable to imagine that closer dwarf galaxies may feed the Milky Way disk via their outflows and tidally stripped material.  One way to differentiate these scenarios would be to study the distribution of the HVC metallicities. At higher redshifts  \hi\ absorbers with $16\le \log N($\hi$)\le 18.5 $ that trace the circumgalactic gas of galaxies (see \S\ref{s-lls}) are sometimes observed at a metallicity of about 2\% solar or less \citep[][N. Lehner et al. 2012, in preparation]{tripp05,ribaudo11}, and these may have originated from cold flow accretion \citep{ribaudo11}.  HVCs with metallicity $<2\%$ solar \citep[like perhaps complex A, see][]{wakker01} could be candidates for cold stream accretion. On the other hand, for $>20\%$ solar metallicity (like complex C), the gas has been polluted to levels as observed in the LMC outflows or Stream, which could be a signature of interaction between nearby dwarf galaxies and the Milky Way or of pollution of more metal poor gas by fountain or thick disk gas. 

There are also HVCs with solar-like metallicity, suggestive of galactic fountain flows \citep[e.g.,][]{shapiro76,bregman80,fraternali06}. For example,  \citet{zech08} derived a super-solar metallicity for two of these HVCs toward the Galactic center. This region is essentially devoid of \hi\ emission at high-velocity (except for complex L), but high-velocity absorption is observed toward several AGN and stars  \citep[][this paper, see Appendix~\ref{s-app}]{sembach03,fox06,keeney06,zech08,bowen08}. The high metallicity and absence of neutral gas are consistent with the HVCs tracing large-scale recycling flows within the central region of the Milky Way as those predicted by Galactic fountain models.

\subsection{Fueling the Milky Way}\label{s-infall}

As the HVCs with $90 \le |v_{\rm LSR}|\la 200$ \km\ are within a few kiloparsecs from the Milky Way irrespective of their \hi\ content, they are also the most likely source of gas for fueling continued star formation in the Milky Way. \citet{collins05} and others argued that the segregation  in positive and negative radial LSR velocity of HVCs with the Galactic coordinates (see Fig.~\ref{f-map}) is consistent with an overall population of infalling clouds that reflect the sense of Galactic rotation, with some peculiar velocities. Although this is not a unique interpretation, our finding that the HVC population drops with decreasing $z$ provides additional support to this conclusion as this drop is naturally explained if HVCs fall through the Milky Way halo onto the disk (see \S\ref{s-fcd}).  A better characterization of the neutral and ionized HVC and IVC populations as a function of $z$ will be needed to know if HVCs actually reach the Galactic disk, but our findings further support that they most likely feed the Galactic disk with new gas.  

Because they both cover a large fraction of the sky and are at distances 5--15 kpc of sun,  the HVCs with \hi\ column density $\log N($\hi$) \la 18.5$ represent a large mass reservoir. The total mass of the predominantly ionized HVCs is $M \approx 1.1 \times 10^8 (d/{\rm 12\,kpc})^2 (f_c/0.5)(Z/0.2Z_{\odot})^{-1}$ M$_{\odot}$, i.e., about $(0.2$--$2) \times 10^8 (Z/0.2Z_{\odot})^{-1}$ M$_{\odot}$ (see LH11). Even though the metallicity can vary from HVC to HVC (see above), this implies that the ionized gas mass is at least as important -- and likely much more important than -- the \hi\ HVCs with  $\log N($\hi$) \ga 18.5$  (excluding the Magellanic Stream), which total about $10^7 M_{\odot}$ \citep{wakker01,putman06,wakker07,wakker08,thom08}. The importance of the ionized HVCs to the total mass budget is also independently demonstrated by the deep H$\alpha$ emission observations. The mass of ionized gas derived from these observations is  similar to the \hi\ mass within the \hi\ contours (i.e., where $\log N($\hi$)>18.5$), but the H$\alpha$ emission contours extend far beyond the \hi\ contours \citep{hill09,barger12}, implying a greater mass of ionized gas. 

While there is still some uncertainty in the fraction of the HVC population that is infalling, the metallicity, and the infall timescales, LH11 conservatively derived an infall rate of $0.4$--$1.4$ M$_\odot$\,yr$^{-1}$. This is enough to balance the present-day total star formation rate ($1.9 \pm 0.4$ M$_\odot$\,yr$^{-1}$) of the Milky Way \citep{chomiuk11} taken into account that the amount of infall needed to sustain star formation may be smaller than the star formation rate  according to chemical evolution models  \citep{chiappini01}  and stellar mass loss models \citep{leitner11}.

\subsection{HVCs in the Local Group and beyond}\label{s-lls}

Most of the HVCs identified so far in the Local Group are found near galaxies. In the Milky Way direct and indirect distance constraints place most of the HVCs and VHVCs within about 5--15 kpc \citep[this paper, LH11][]{wakker01,putman03}, with only the Magellanic Stream possibly being much farther \citep{besla12}.  Around M31 and M33, the \hi\ HVCs are found within about 50 kpc, and many are within 15 kpc \citep{thilker04,westmeier05,westmeier08,putman09}. Deep \hi\ observations have also revealed a tenuous filament of gas between M31 and M33 separated by about 150 kpc, which could be the result from a tidal interaction between these two galaxies  \citep{braun04}.  The HVCs observed toward the LMC probe outflows from the LMC and are also likely near the host galaxy \citep{lehner09}.  HVCs in the Local Group of galaxies, including our Milky Way,  are thus all within 10--50 kpc. While previously it could be argued that purely ionized HVCs could be farther away \citep[e.g.,][]{westmeier08}, our findings show that the ionized and neutral gas in HVCs are at similar distances, at least in the Milky Way halo. Although the sample of galaxies is small, there is therefore no evidence in our local galactic neighbor that ionized or neutral HVCs are found much beyond 1--2 galactic radii, except possibly for the remnants of galaxy interactions.\footnote{We also emphasize that while in the Milky Way it might be plausible that  a population of HVCs may be missed if these have low velocities, this is not the case for M31 or M33 where ``HVCs" with any velocities (included systemic velocities of M31 and M33) can be clearly spatially separated from the galaxy.} 

If we move out from our local neighborhood, attempts have been made to connect QSO absorbers -- in particularly  the optically thick absorbers (a.k.a., the Lyman limit systems, LLSs) -- with the HVCs  \citep[e.g.,][]{charlton00,richter09}. While some of the LLSs are likely higher redshift analogs of HVCs, we emphasize that the distances of HVCs place them close to galaxies. HVCs also probe ``absorbers" with virtually any $N($\hi$)$ and therefore likely include the population of both weak and strong \mgii\ absorbers. 

\section{Summary}\label{s-con}
In this work we have built on the results presented by LH11 and reported on the covering factors and distances of the HVCs, as well as the associations of  HVCs observed in \hi\ 21-cm emission with those found in absorption.   We emphasize that the UV absorption diagnostics are much more sensitive to low column density gas  than \hi\ emission (or optical absorption diagnostics such as \caii; e.g., Richter et al. 2011). This  allows us to probe clouds over several orders of magnitude in \hi\ column density, from $N($\hi$)<10^{15}$ cm$^{-2}$ to $N($\hi$)>10^{20}$ cm$^{-2}$. We are also able to probe a wide range of ionization conditions and densities because we use several species in different ionization stages (e.g., \oi, \cii, \civ, \siii, \siiii, \siiv). We summarize the results of our study as follows:

\begin{enumerate}
\item With a carefully-selected, large sample of AGN that cover the sky at $|b|\ga 20\degr$, we find that the covering factor of the HVCs with $|v_{\rm LSR}|\ge 90$ \km\ of the entire Galactic sky is $68\% \pm 4\%$.  About $74\%$ of the HVC directions have $N($\hi$)<3\times 10^{18}$ cm$^{-2}$ and $46\%$ have $N($\hi$)<8\times 10^{17}$ cm$^{-2}$ from a comparison of our results with 21-cm \hi\ emission surveys. 

\item The full sample of AGN  and a sample with more uniform and higher S/N data gives the same covering factor. HVCs with a total H column density   $N({\rm H})< 10^{17} (Z/Z_{{\odot}})^{-1}$ cm$^{-2}$ are unlikely to play an important role in the total mass of the HVCs. However, HVCs with $N($\hi$)\ll 10^{17}$ cm$^{-2}$ are not rare and have still significant mass in view of the detections of metal ion absorption.  

\item With the distance, position, and velocity information, we unambiguously associate the denser (\hi\ emission selected) and more diffuse (UV absorption selected) regions of HVCs. We argue that most of the predominantly neutral HVC complexes have ionized envelopes that extend beyond the \hi\ contours and that the predominantly ionized HVCs contain at least as much mass as the \hi\ component. However, there are also large regions of Galactic sky that are filled with ionized high-velocity gas with little evidence for \hi\ counterparts nearby. One such region is found toward the Galactic center and may be generated by Galactic outflows. 

\item The covering factors of HVCs with $90 \le |v_{\rm LSR}|\la 170$ \km\ determined from the AGN and LH11 stellar samples are very similar. This confirms that HVCs are within 5--15 kpc of the sun. The HVCs are therefore flows of gas in the inner Milky Way halo. Our new results also show that the covering factor  of HVCs drops with decreasing $z$, which is consistent with theoretical predictions of HVCs falling onto the Milky Way disk.  The HVCs are far enough to have a substantial mass ($M\propto d^2$), but are also near enough to be able to reach the disk. They are therefore the most likely source of gas required to support continued star formation in the Milky Way.

\item While there is no large disparity in the covering factor of the HVCs between the two Galactic hemispheres, the situation is very different for the VHVCs (HVCs with $|v_{\rm LSR}|>170$ \km). At $b>0\degr$, the covering of the VHVC is only 13\%, but at $b<0\degr$, 46\% of the Galactic sky is covered by VHVCs. Most of the VHVCs are associated with the Magellanic Stream at $b<0\degr$; a large fraction also appears to be related to the leading arm of the Stream (possibly even at higher positive latitude than previously thought). Other VHVCs with $170 \la |v_{\rm LSR}|\la 210$ \km\ are affiliated with complex A, C,  and the anti-center complex. 

\item All these elements strongly suggest that in our Local Group of galaxies there is no evidence that HVCs are found much beyond 50 kpc from their host galaxy, expect for structures that are directly related to galaxy interaction (like the Magellanic Stream). However, compact VHVCs may have been mostly missed along  pencil-beam sightlines; these compact HVCs are the only remaining HVC candidates for dark matter mini-halos. 

\end{enumerate}

\section*{Acknowledgments}
We thank Bart Wakker for useful discussions. Support for this research was provided by NASA through grants  HST-GO-11592.01-A, HST-GO-11598.01, and HST-GO-11741.01-A  from the Space Telescope Science Institute, which is operated by the Association of Universities for Research in Astronomy, Incorporated, under NASA contract NAS5-26555. All of the data presented in this paper were obtained from the Multi-mission Archive at the Space Telescope Science Institute (MAST). STScI is operated by the Association of Universities for Research in Astronomy, Inc., under NASA contract NAS5-26555. This research has made use of the NASA Astrophysics Data System Abstract Service and the SIMBAD database, operated at CDS, Strasbourg, France.

\appendix

\section{Distance of HVC complexes}\label{s-app}

We discuss below the connection or absence of connection between the HVCs seen in absorption toward the stars from the LH11 sample (and one additional star HD195455) and  in \hi\ emission.

-- {\it NGC5904-ZNG1:}\ No high-velocity \hi\ emission is observed along this sightline, which is not surprising as \citet{zech08} derived $\log N($\hi$) = 16.5$ using the Lyman series. The HVCs at $v_{\rm LSR} = -140,-120 $ \km\ have similar velocities as those observed in complex L seen several degrees away \citep[for more information and the spectra, see][]{zech08}. Complex L and this star are in the general direction of the Galactic center where lie also the stars NGC6723-III60 (negative latitude) and PG1323-086 (positive latitude). Toward these stars, high-velocity absorption is observed at about $-90$ \km\ (see Fig.~\ref{f-spec} and Table~\ref{t-dist}), but they are not close to any known \hi\ HVC complex. Similarly, AGN and deep \hi\ observations also show additional HVCs at both negative and positive velocities toward the Galactic center \citep{keeney06,lockman02}.  It would be interesting to map out the entire region of the Galaxy center with deep \hi\ and H$\alpha$ observations at $|v_{\rm LSR}|>90$ \km\ to determine the extent of the ionized gas in this region and its possible connection with complex L. The HVCs observed toward NGC5904-ZNG1 have a very well determined metallicity based on \oi\ and \hi\ absorption, which is super-solar ($[$O/H$]=+0.22 \pm 0.10$). (Another test for the connection or not of these clouds with complex L would be to measure the metallicity of complex L). In similar directions, \ovi\ absorption is also detected at these positive and negative velocities closer to the plane \citep[e.g., toward HD168941, see][]{bowen08}, so these HVCs could be affiliated with flows originating in the disk. The super-solar metallicity, large column of predominantly ionized gas seen in the direction of the inner Galaxy, presence of both positive and negative velocity may suggest that they participate to a galactic fountain fed by outflows occurring in the central regions of the Milky Way. 

-- {\it M13-Barnard29:}\ An HVC is observed in absorption at $-121$ \km\ \citep[see the spectra in][]{welsh11}. The sightline passes near to (but outside of) the \hi\ contours of complex C and no LAB \hi\ emission is observed at $-121$ \km. However, the absorption velocity is similar to velocities observed in the nearby \hi\ emission from complex C. The distance of complex C was derived to be at $d= 10 \pm 2.5$ kpc (Thom et al. 2008, and see also Wakker et al. 2007). While the distance of M13-Barnard29, 7.1 kpc, is slightly outside this distance bracket, it is not inconsistent, especially taken into account a 10\% error on the distance of M13-Barnard29, possible variation in the distances of complex C with position, and similarity in the velocities.  Therefore this sightline probably traces the extended ionized envelope of complex C. We therefore disagree with the conclusions of \citet{welsh11} who argued that the distance of this star was too small for the gas to be associated with complex C. 

-- {\it HD195455:}\ This star lies in the general direction of the IVC complex gp, which spans LSR velocities between $+55$ and $+90$ \km. The HVC observed at $+95$  \km\ toward this direction could be a higher velocity extended ionized component of this complex. The IVC Complex gp component is also observed  at $+70$ \km\ along this sightline (see Fig.~\ref{f-spechd}).  The distance of HD195455 is consistent with the bracketed distance of  complex gp between 1.8 and 3.8 kpc \citep{wakker08}.

-- {\it HS1914+7139:}\ Two HVCs were detected along this sightline at $v_{\rm LSR} = -118, -175$ \km, which are associated with the Outer Arm and/or complex C \citep[see for more information and the spectra in][and also discussion in Tripp \& Song 2012]{lehner10}. LAB \hi\ emission was detected only in the $-112$ \km\ component. 

-- {\it PG0122+214:}\ This line of sight has also two HVCs detected in absorption at $-91$ and $-160$ \km\ (see Fig.~\ref{f-spec}) and passes near the anti-center HVC complexes, more specifically near the Cloud WW507 and the Cohen stream, where similar velocities are observed. Again no high-velocity \hi\ emission is observed along this sightline. The distance of PG0122+214 (9.6 kpc) is consistent with the bracket distance of 5--11.7 kpc derived by \citet{wakker08}. However, the higher velocity component was not observed in the stellar spectra of \citet{wakker08}, so this star places a new upper limit to the distance of the higher velocity gas observed in this complex. 

-- {\it PG0832+675:}\ This line of sight differs with the others in the sense that \hi\ emission is observed with the peak emission  centered at $-145$ \km, which is different from the observed absorption centered at $-123$ \km. In Fig.~\ref{f-pg0832}, we show the LAB \hi\ 21-cm emission and the normalized spectra of \oi, \cii, \aliii, \siii, \sii, and \feii. The central velocities of the low and intermediate velocity components seen in \sii\ absorption and \hi\ emission are the same, and therefore the difference between the absorption and emission  at high velocities is real. Although the UV spectrum is complicated by many stellar features, the component at $-123$ \km\ is very likely  interstellar because the velocity is the same for different species (\cii, \oi, \nni, \siii, \feii, although we note that \oi\ and \nni\ are likely contaminated at $-123$ \km) and different from the star ($-70$ \km), and because the column densities derived from different transitions of the same species give the same values ($\log N($\feii$)=12.65 \pm 0.02$ based on the $\lambda$2344, 2382 transitions and $\log N($\aliii$)=12.55 \pm 0.05$ based on the $\lambda$1854, 1862 transition,  and $\log N($\siii$)=13.00 \pm 0.08$ based on the $\lambda$1193, 1260, 1304 transitions). 

PG0832+675 lies in the direction of complex A within the \hi\ contour, so it is not surprising to see \hi\ emission in this direction. The star is likely at the front edge of complex A, which would explain why the \oi\ is so weak at $-145$ \km\ (the metallicity of complex A --B.P. Wakker, private communication 2011 -- would imply a far stronger absorption). \citet{wakker01} summarized a distance bracket for this complex of 4.0--9.9 kpc \citep[see][]{wakker96,vanwoerden99}. Given that we see little \oi, \cii, \siii, and \feii\ absorption at $-145$ \km\ toward PG0832+675 compared to the \hi\ column density at the same velocity, we conclude that most of the \hi\ lies behind the star. This implies a distance bracket for the predominantly neutral gas of complex A of 8.1--9.9 kpc. The $-123$ \km\ component may be the ionized envelope of complex A or an unrelated foreground cloud. The upper limit for its distance, in either case, is thus 8.1 kpc. While \citet{barger12} observed  H$\alpha$ emission in complex A at  LSR velocities ranging from $-220$ to $-110$ \km,  there is no strong evidence of H$\alpha$ emission at $-123$ \km\ over a $1\degr$ beam toward this star (K. Barger 2011, private communication). This does not rule out that the $-123$ \km\ component is part of complex A. 

-- {\it PG1002+506:}\  The \hi\ HVC WW503 is about $2.5\degr$ away from this sight line with $v_{\rm LSR}=-105$ \km\ (B.P. Wakker, private communication 2011), similar to the velocity observed in absorption. Our observations provide a new upper limit, $<13.9$ kpc, on the distance to the ionized component of this cloud. This strongly suggests that most of the HVCs in the anti-center region at $b<0\degr$ are within about 10 kpc \citep[see][]{wakker08}. On the other hand, no \hi\ emission is observed near this direction at $v_{\rm LSR} = +101$ \km, indicating the presence of a positive velocity HVC that is not associated with a \hi\ complex at a distance $d<13.9$ kpc. 

-- {\it PG0855+294:}\ Several small high-velocity \hi\ clouds are observed in this general direction. This sightline passes $0.5\degr$ away from HVC WW113 seen in \hi\ emission at $v_{\rm LSR} = -90$ \km (B.P. Wakker, private communication 2011),  setting an upper limit of 6.5 kpc for the extended ionized component of these clouds. 

-- {\it PG0914+001:}\ This sightline is near complex WA ($v_{\rm LSR} = +140$--$+170$ \km) and WB ($v_{\rm LSR} = +80$--$+110$ \km), and absorption features at $+110$ and $+170$ \km\ are observed in the spectrum of this star (for more information and the spectra, see LH11).  This places an upper limit of  16 kpc  for the extended ionized envelope of these HVCs. For the WB complex, the HVCs could be closer as \citet{thom06} derived a bracket distance of 7.5--11.1kpc. 

-- {\it EC10500-1358:}\ This sightline is also in the general direction of complex WA, but the velocity seen in absorption ($v_{\rm LSR}=+97$ \km, see the spectra in Fig.~\ref{f-spec}) is smaller. It is, however, about $0.5 \degr$ from HVC WW95 at $v_{\rm LSR}=+100$ \km, setting a distance limit of 5.2 kpc for this cloud. 

-- {\it NGC104-UIT14:}\ This sightline shows no HVC absorption, although it has strong \hi\ 21-cm emission observed at $+112$ \km\ , with a \hi\ column density $\log N($\hi$) = 20.39$. This sightline passes in the southern part of the WE complex \citep{morras00,wakker01}, which is a patchy ensemble of HVCs at similar velocities rather than a large continuous structure like complex C.  The limit from \oi\ gives $\log N($\oi$)< 13.8$ at $3\sigma$. It is unlikely that the metallicity in this complex is $<0.5 \times 10^{-3} Z_{\odot}$, implying that the star sets a lower limit on the distance of WE, and hence using the upper limit from \citet{sembach91}, the WE complex is at $4.5<d<12.8$ kpc.

\bsp

\label{lastpage}

\end{document}